\documentclass{aa}

\usepackage{graphicx}
\usepackage{txfonts}
\usepackage{booktabs}
\usepackage{tabularx}
\usepackage{hyperref}
\usepackage{siunitx}
\def\cch{C$_2$H}
\def\gal{NGC~253}

\begin{document}

   \title{The distribution and origin of \cch~in \gal~from ALCHEMI}

   \subtitle{}

   \author{
        J. Holdship\inst{\ref{inst.Leiden},\ref{inst.UCL}}
        \and S. Viti\inst{\ref{inst.Leiden},\ref{inst.UCL}}
        \and S. Mart\'in \inst{\ref{inst.ESOChile},\ref{inst.JAO}}
        \and N. Harada \inst{\ref{inst.NAOJ},\ref{inst.ASIAA},\ref{inst.SOKENDAI}}
        \and J. Mangum \inst{\ref{inst.NRAOCV}}
        \and K. Sakamoto \inst{\ref{inst.ASIAA}}
        \and S. Muller \inst{\ref{inst.ONSALA}}
        \and K. Tanaka \inst{\ref{inst.KeioUniversity}}
        \and Y. Yoshimura  \inst{\ref{inst.UTokio}}
        \and K. Nakanishi \inst{\ref{inst.NAOJ},\ref{inst.SOKENDAI}}
        \and R. Herrero-Illana \inst{\ref{inst.ESOChile},\ref{inst.ICECSIC}}
        \and S. M\"uhle \inst{\ref{inst.UBonn}}
        \and R. Aladro \inst{\ref{inst.MaxPlanck}}
        \and L. Colzi \inst{\ref{inst.MadridCAB},\ref{inst.INAF}}
        \and K. L. Emig \inst{\ref{inst.NRAOCV},}\thanks{Jansky Fellow of the National Radio Astronomy Observatory}
        \and S. Garc\'ia-Burillo\inst{\ref{inst.OAN}}
        \and C. Henkel \inst{\ref{inst.MaxPlanck},\ref{inst.Abdulaziz}}
        \and P. Humire
        \inst{\ref{inst.MaxPlanck}}
        \and D. S. Meier \inst{\ref{inst.NMIMT},\ref{inst.NRAOSocorro}}
        \and V. M. Rivilla \inst{\ref{inst.MadridCAB},\ref{inst.INAF}}
        \and P. van der Werf
        \inst{\ref{inst.Leiden}}
        }
 \institute{
\label{inst.Leiden}Leiden Observatory, Leiden University, PO Box 9513, NL - 2300 RA Leiden, The Netherlands\email{holdship@strw.leidenuniv.nl}
\and\label{inst.UCL}Department of Physics and Astronomy, University College London, Gower Street, London WC1E6BT, UK
\and\label{inst.ESOChile}European Southern Observatory, Alonso de C\'ordova, 3107, Vitacura, Santiago 763-0355, Chile
\and\label{inst.JAO}Joint ALMA Observatory, Alonso de C\'ordova, 3107, Vitacura, Santiago 763-0355, Chile
\and\label{inst.NAOJ}National Astronomical Observatory of Japan, 2-21-1 Osawa, Mitaka, Tokyo 181-8588, Japan
\and\label{inst.ASIAA}Institute of Astronomy and Astrophysics, Academia Sinica, 11F of AS/NTU Astronomy-Mathematics Building, No.1, Sec. 4, Roosevelt Rd, Taipei 10617, Taiwan   
\and\label{inst.SOKENDAI}Department of Astronomy, School of Science, The Graduate University for Advanced Studies (SOKENDAI), 2-21-1 Osawa, Mitaka, Tokyo, 181-1855 Japan
\and\label{inst.NRAOCV}National Radio Astronomy Observatory, 520 Edgemont Road, Charlottesville, VA 22903-2475, USA
\and\label{inst.ONSALA}Department of Space, Earth and Environment, Chalmers University of Technology, Onsala Space Observatory, SE-43992 Onsala, Sweden
\and\label{inst.KeioUniversity}Department of Physics, Faculty of Science and Technology, Keio University, 3-14-1 Hiyoshi, Yokohama, Kanagawa 223--8522 Japan
\and\label{inst.UTokio}Institute of Astronomy, Graduate School of Science, The University of Tokyo, 2-21-1 Osawa, Mitaka, Tokyo 181-0015, Japan
\and\label{inst.ICECSIC}Institute of Space Sciences (ICE, CSIC), Campus UAB, Carrer de Magrans, E-08193 Barcelona, Spain
\and\label{inst.UBonn}Argelander-Institut f\"ur Astronomie, Universit\"at Bonn, Auf dem H\"ugel 71, D-53121 Bonn, Germany
\and\label{inst.OAN}Observatorio Astron\'omico  Nacional (OAN-IGN), Observatorio de Madrid, Alfonso XII, 3, 28014-Madrid, Spain
\and\label{inst.MaxPlanck} Max-Planck-Institut f\"{u}r Radioastronomie,
 Auf dem H\"{u}gel 69, D-53121 Bonn, Germany
\and\label{inst.MadridCAB} Centro de Astrobiolog\'ia (CSIC-INTA), Ctra. de Ajalvir Km. 4, Torrej\'on de Ardoz, 28850 Madrid, Spain
\and\label{inst.INAF} INAF-Osservatorio Astrofisico di Arcetri, Largo Enrico Fermi 5, 50125, Florence, Italy
\and\label{inst.Abdulaziz}Astron. Dept., Faculty of Science, King Abdulaziz University, P.O. Box 80203, Jeddah 21589, Saudi Arabia
%
\and\label{inst.NMIMT}New Mexico Institute of Mining and Technology, 801 Leroy Place, Socorro, NM 87801, USA
\and\label{inst.NRAOSocorro}National Radio Astronomy Observatory, PO Box O, 1003 Lopezville Road, Socorro, NM 87801, USA
         }
   \date{Received; accepted}

 
  \abstract
   {Observations of chemical species can provide insights into the physical conditions of the emitting gas however it is important to understand how their abundances and excitation vary within different heating environments. \cch~is a molecule typically found in PDR regions of our own Galaxy but there is evidence to suggest it also traces other regions undergoing energetic processing in extragalactic environments.}
   {As part of the ALCHEMI ALMA large program, we map the emission of \cch~in the central molecular zone of the nearby starburst galaxy NGC 253 at 1.6 " (28 pc) resolution and characterize it to understand its chemical origins.}
   {We used spectral modeling of the N=1-0 through N=4-3 rotational transitions of \cch~ to derive the \cch~column densities towards the dense clouds in NGC 253. We then use chemical modeling, including photodissociation region (PDR), dense cloud, and shock models to investigate the chemical processes and physical conditions that are producing the molecular emission.}
   {We find high \cch{} column densities of $\sim$\SI{e15}{\per\centi\metre\squared} detected towards the dense regions of \gal. We further find that these column densities cannot be reproduced if it is assumed that the emission arises from the PDR regions at the edge of the clouds. Instead, we find that the \cch~abundance remains high even in the high visual extinction interior of these clouds and that this is most likely caused by a high cosmic-ray ionization rate.}
   {}

   \keywords{Astrochemistry --
                 Galaxies: abundances --
                 Galaxies: starburst --
                 Galaxies: individual: NGC253
               }

   \maketitle
%

\section{Introduction}
Observations of chemical species provide a wealth of information about the physical conditions of astrophysical objects. However, in order to get the most information from a particular observation, it is important to understand the chemistry and excitation of molecules and related atomic species under different conditions. Whether the species under study is a classical dense gas tracer like CS \citep[eg.,][]{vanderTak2000StructureStars} or more characteristic of photon-dominated regions (PDRs) such as C$^+$, many of the most useful tracers are those whose relationships to physical parameters of interest are well understood.\par
Within the Milky Way, ethynyl (\cch) is a typical photodissociation region (PDR) tracer as its chemistry is strongly linked to C$^+$ \citep{Meier2005Spatially342}. It is observed at low fractional abundances ($\sim$\num{e-10}) in dense clouds \citep{Wootten1980DetectionClouds,Watt1988CCHClouds.} and enhanced to $\sim$\num{e-8} in low density gas \citep{Turner1999TheProcesses,Lucas2000ComparativeC_3H_2}, where UV photons can penetrate and ionize C which is a key reactant in a chain of reactions that form \cch. Whilst \cch~is also abundant in these environments in extragalactic objects \citep[eg.,][]{Aladro2011A82,Aladro2015LambdaGalaxies}, there is an indication that it may be tracing more than just PDR regions in those extragalactic environments \citep{Garcia-Burillo2017}. \par
A previous work focused on the Seyfert galaxy NGC~1068 \citep{Garcia-Burillo2017} indicated that \cch~could be a tracer of the interface between high energy outflows and the ambient gas in a galaxy. The starburst ring of this galaxy showed a typical \cch~abundance ($\sim$ \num{e-8}) similar to that found in diffuse clouds in Galactic environments. However, a larger, more extended component of gas had an abundance of $\sim$\num{e-6} in the region between the molecular disk and ionized gas outflow. Chemical modeling efforts showed this larger abundance was only consistent with the short, early stages of models where some energetic process such as high cosmic ray ionization or shocks enhanced the \cch~abundance. It was theorized in that work that the interface between an outflow and ambient gas would create an environment where unprocessed gas would continually be replenished and subjected to energetic processing. This could create a pseudo-steady state where the \cch~abundance is stable at the high value found in the early stages of simple shock models.\par
It is possible therefore that \cch~is a good tracer of molecular gas that is undergoing energetic processing and, more specifically, of the interface between active galactic nuclei (AGN) driven outflows and their environment. This would make it a useful tool for extragalactic observations. However, this must be tested against other galaxies. Moreover, only one multiplet of \cch~was observed towards NGC~1068 and so additional transitions must be observed to assess excitation conditions and thus give a clearer picture.\par
\gal~is a prototypical starburst galaxy that hosts several large ($\sim$ 30 pc), dense ($\sim$\SI{e5}{\per\centi\metre\cubed}) clouds which are well studied \citep[e.g.,][]{Sakamoto2011,Leroy2018}. Though similar in size to giant molecular clouds (GMCs) in the Milky Way, they are orders of magnitude more massive and have higher velocity dispersions \citep{Leroy2015}. \gal~ was the chosen target of the ALCHEMI ALMA large program \citep{Martin2021ALCHEMI}, which aimed at obtaining the most complete extragalactic molecular inventory in the central molecular zone of a starburst galaxy at a spatial resolution of tens of parsecs. \gal~also has a starburst driven outflow which can be observed at X-ray wavelengths \citep{Dahlem1998AnData} and in molecular emission \citep{Bolatto2013SuppressionWind,Krieger2019}. This outflow may lead to a chemical environment similar to the AGN driven outflow of NGC~1068. \par
In summary, \gal~presents an interesting test case for the scenario proposed and modeled for NGC~1068. The star-forming regions of \gal~should present \cch~emission similar to that found in the starburst ring of NGC~1068. More importantly, if enhanced \cch~abundances are observed along the outflow then the chemical origin must be due to some commonality between the outflows from both galaxies. Alternatively, if enhanced abundances are not observed, this would indicate a different chemical composition between starburst-driven outflows and AGN-driven outflows.\par
To investigate this, images of multiple transitions of \cch~in the central molecular zone (CMZ) of \gal, as well as calculations of the column density and fractional abundance are presented. In Section~\ref{sec:obs}, the observations and ALCHEMI large program are described. In Section~\ref{sec:results}, the \cch~emission distribution and column densities are discussed. In Section~\ref{sec:chemicalmodels}, average fractional abundances are interpreted through chemical models.\par

\section{Observations \& image processing}
\label{sec:obs}
\subsection{Observations}
The analysis presented in this article makes use of the data collected by the ALMA ALCHEMI large program, which is an unbiased spectral survey of \gal~covering the full ALMA spectral bands 3 through 7. The ALCHEMI program and all observational details are described extensively in \citet{Martin2021ALCHEMI}.  A summary of ALCHEMI survey details pertinent to the \cch~data presented in this article is given here.\par
\gal~was observed toward a nominal phase center of $\alpha=00^h47^m33.26^s$, $\delta=-25^\circ17'17.7''$ (ICRS).
Observations were configured to cover a common rectangular area of $50\arcsec\times20\arcsec$ with a position angle of $65^\circ$ (East of North).  The spectral configuration for the ALCHEMI Survey of \gal~was configured with 47 receiver tunings; each comprising four 1.875 GHz spectral windows. All tunings were imaged to a shared beam size of 1\farcs6 and the maximum recoverable scale was 15\arcsec \footnote{The maximum recoverable scale corresponds to the size of the largest structure that can be observed with a given array configuration of an interferometer. In practice, it can be estimated as $\lambda/B_{min}$, where $\lambda$ is the wavelength and $B_{min}$ is the length of the shortest projected baseline during the observations. [See also \url{https://almascience.nrao.edu/documents-and-tools/cycle8/alma-technical-handbook/view}]}. At a distance of 3.5~Mpc \citep{Rekola2005DistanceFunction}, these correspond to linear scales of 28 and 250~pc, respectively. Within the ALCHEMI frequency coverage, four multiplets of \cch~transitions namely N=1-0, 2-1, 3-2, and 4-3 were observed. This includes 39 hyperfine transitions across ALMA bands 3, 5, 6, and 7 and these are listed in Table~\ref{table:cchlines}.\par
\begin{table}[]
    \centering
    \caption{\cch~transitions covered by the ALCHEMI large program's observations taken from JPL \citep{pickett1998}.}
    \begin{tabular}{ccccc}
        \toprule
        {\bf Frequency / GHz} & {\bf Quantum Numbers} & {\bf log$_{10}$(A$_{ij}$)} & {\bf E$_U$ } \\
        GHz & & log$_{10}$(s$^{-1}$) & K \\
        \midrule
        & {\bf N=1-0} & & \\
        \midrule

        87.284 & J=3/2-1/2,F=1-1 & -6.43 & 4.2 \\
        87.317 & J=3/2-1/2,F=2-1 & -5.66 & 4.2 \\
        87.329 & J=3/2-1/2,F=1-0 & -5.74 & 4.2 \\
        87.402 & J=1/2-1/2,F=1-1 & -5.74 & 4.2 \\
        87.407 & J=1/2-1/2,F=0-1 & -5.65 & 4.2 \\
        87.447 & J=1/2-1/2,F=1-0 & -6.42 & 4.2 \\
        \toprule
        & {\bf N=2-1} & & \\
        \midrule
        174.663 & J=5/2-3/2,F=3-2 & -4.67 & 12.6 \\
        174.668 & J=5/2-3/2,F=2-1 & -4.71 & 12.6 \\
        174.722 & J=3/2-1/2,F=2-1 & -4.78 & 12.6 \\
        174.728 & J=3/2-1/2,F=1-0 & -4.93 & 12.6 \\
        174.733 & J=3/2-1/2,F=1-1 & -5.13 & 12.6 \\
        174.807 & J=3/2-3/2,F=2-2 & -5.41 & 12.6 \\
        \toprule
        & {\bf N=3-2} & &\\
        \midrule
        262.004 & J=7/2-5/2,F=4-3 & -4.11 & 25.2 \\
        262.006 & J=7/2-5/2,F=3-2 & -4.13 & 25.2 \\
        262.064 & J=5/2-3/2,F=3-2 & -4.15 & 25.2 \\
        262.067 & J=5/2-3/2,F=2-1 & -4.19 & 25.2 \\
        262.078 & J=5/2-3/2,F=2-2 & -5.06 & 25.2 \\
        262.208 & J=5/2-5/2,F=3-3 & -5.24 & 25.2 \\
        \toprule
        & {\bf N=4-3} & &\\
        \midrule
        349.337 & J=9/2-7/2,F=5-4 & -3.72 & 41.9 \\
        349.338 & J=9/2-7/2,F=4-3 & -3.73 & 41.9 \\
        349.398 & J=7/2-5/2,F=4-3 & -3.74 & 41.9 \\
        349.400 & J=7/2-5/2,F=3-2 & -3.76 & 41.9 \\
        \bottomrule
    \end{tabular}
    \label{table:cchlines}
\end{table}
The datacubes used in this work have been made available as part of the public release of the ALCHEMI data. The FITS files associated with the B3a, B5d, B6g, and B7o science goals were used for this work each covering a single multiplet from N=1-0 to N=4-3.
\subsection{Data processing}
We loaded and processed the FITS files using Astropy and Spectral Cube \citep{Robitaille2013,Price-Whelan2018}. Standard routines from those packages were then used to produce moment 0 maps. The velocity ranges for those maps are given in Table~\ref{table:limits} and were chosen so that they contain all the \cch~lines but excluded any strong lines from other species that may contaminate the maps. This was informed by local thermodynamic equilibrium (LTE) modeling of the ALCHEMI project's ALMA compact array data of the region \citep{Martin2021ALCHEMI} that provided both the velocity ranges that would cover all expected \cch~emission and  the line intensities of possible contaminating lines. Only the N=3-2 transition was likely to be contaminated and, as a result, we chose a velocity interval that excluded a hyperfine component with frequency 261.834 GHz in order to remove the $7_6-6_5$ line of SO. The spectral modeling performed in this work (Section~\ref{sec:gmc-models}) predicted negligible emission from the omitted \cch~line and verified that the chosen velocity ranges contained all other \cch~emission.\par
\begin{table}
    \centering
    \caption{Velocity limits used to produce the moment 0 maps for each \cch~multiplet.}
    \begin{tabular}{ccc}
    \toprule
        {\bf Multiplet} & {\bf Transition Frequency}  & {\bf Velocity Range} \\
        & {\bf GHz} & {\bf \si{\kilo\metre\per\second}}\\
        \midrule
        N=1 -- 0 & 87.407 & $-185$ - 370\\
        N=2 -- 1 & 174.663 & $-220$ - 380\\
        N=3 -- 2 & 262.004 & $-185$ - 310\\
        N=4 -- 3 & 349.337 & $-185$ - 415 \\
    \bottomrule
    \end{tabular}
    \tablefoot{Velocities are given according to the radio convention using the frequency of the highest A$_{ij}$ in each multiplet. The wide ranges account for multiple components rather than being determined by line width.}
    \label{table:limits}
\end{table}
We also extracted spectra using Spectral Cube and used the package's unit conversion to convert the extracted spectra to units of Kelvin. We estimated the noise value for any given spectrum from the mean difference between the median value and all channels lower than the median. This is effectively taking the variance whilst excluding channels above the average as they are "contaminated" by signal. Where the spectra were extracted from multiple pixels, such as in the GMCs (Section~\ref{sec:distribution}), a median stack of the individual pixel spectra was taken.
\subsection{Spectral modeling}
 \cch~is a molecule with hyperfine structure that, together with the broad velocity distributions in the data, produces spectra with large numbers of blended lines. Furthermore, the complex dynamic situation in \gal~gives rise to multiple, blended emitting components \citep{Krieger2020a}. As such, a spectral model is required to disentangle the emission.\par
In this work, we used SpectralRadex\footnote{\url{https://spectralradex.readthedocs.io}} to model the spectra. This is a Python package that creates model spectra from RADEX using a formalism described in Appendix~\ref{app:spectralradex}. Briefly, it contains a wrapper to run RADEX \citep{VanderTak2007} in order to obtain the excitation temperature and the optical depth at the line centre of every modeled transition. The package then produces model spectra by assuming that each transition follows a Gaussian line profile described by the central velocity and FWHM. It has been benchmarked against the RADEX based spectral model in CASSIS\footnote{\url{http://cassis.irap.omp.eu/}} \citep{Vastel2015} for a wide variety of test cases and each best fit presented in this work has been replicated in CASSIS. Collisional rates between \cch~and ortho and para H$_2$ were used \citep{Dagdigian2018HyperfinePara-H2} to produce the RADEX outputs assuming an ortho to para ratio of 3:1. These cover a temperature range of 10 to \SI{300}{\kelvin} and were taken from the LAMDA database\footnote{\url{https://home.strw.leidenuniv.nl/~moldata/}} \citep{Schoier2005}.\par
\section{Results}
\label{sec:results}
\subsection{Distribution of C$_2$H}
\label{sec:distribution}
\begin{figure*}
	\includegraphics[width=\textwidth]{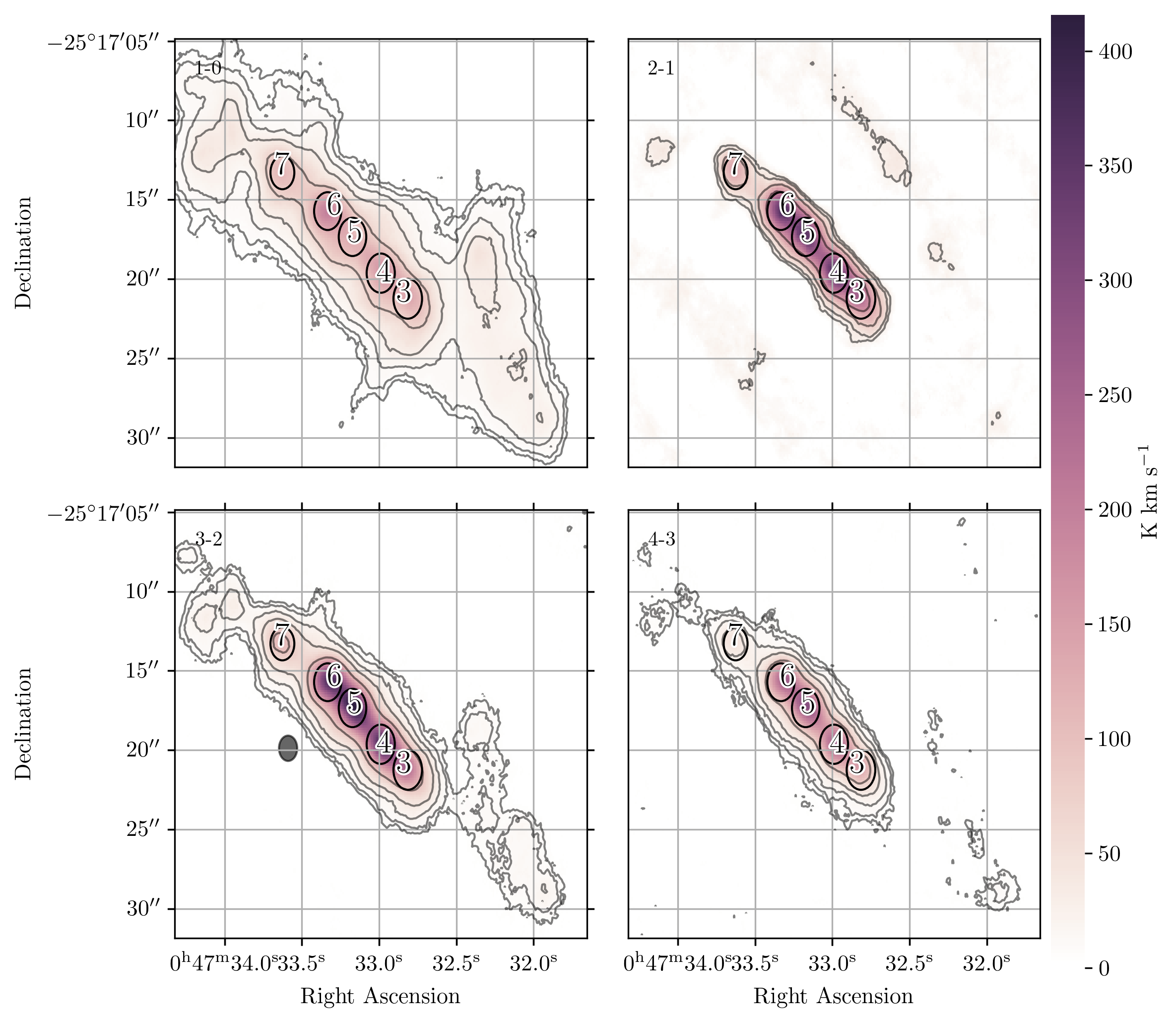}
    \caption{Integrated emission maps of \cch~in \gal. The beam size for all panels is 1.6'' and is shown in the bottom left panel. Since the beam size is shared, the wider spatial distribution of the N=1-0 emission is not caused by a beam size effect. Contours at the 3, 5, 10, 20 and 50$\sigma$ level are drawn in black. The noise levels in order of increasing N are 1.4, 8.5, 2.0 and \SI{1.6}{\kelvin\kilo\metre\per\second}. Each ellipse shows the position and size of the Gaussian fits to the four individual clumps described in section~\ref{sec:distribution}. The numbers indicate the position of the GMCs in \citet{Leroy2018}}
    \label{fig:multimap}
\end{figure*}

\begin{figure}
    \centering
    \includegraphics[width=0.45\textwidth]{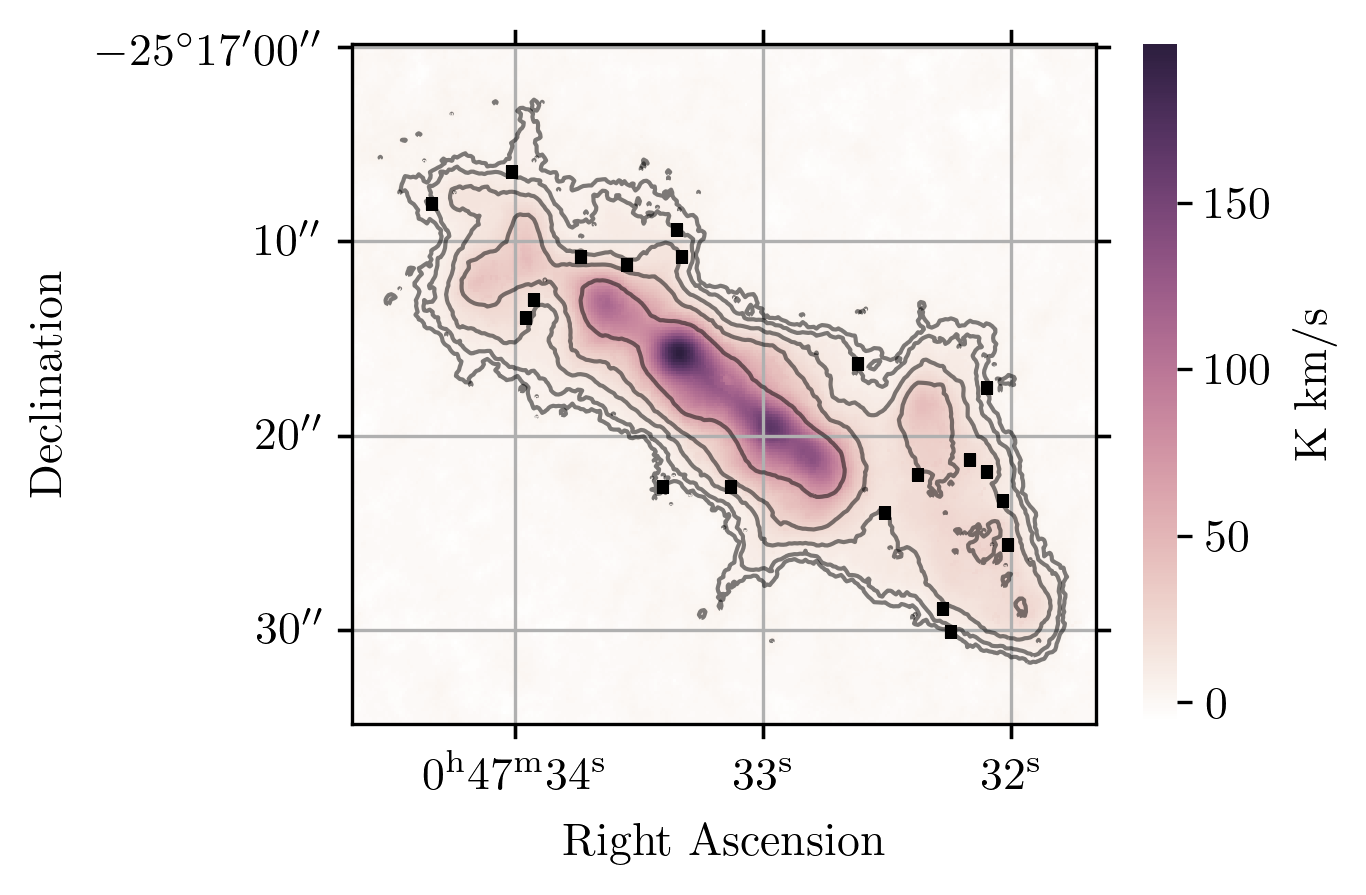}
    \caption{Integrated emission map of the \cch~N=1-0 multiplet as in Figure~\ref{fig:multimap} but showing the full extent of the emission. Contours are shown at the 3, 5, 10, 20 and 50 $\sigma$ levels for noise level of $\sigma$=\SI{1.5}{\kelvin\kilo\metre\per\second}. Black squares indicate the positions sampled for spectra in Figure~\ref{fig:extendedspectra}.}
    \label{fig:1-0map}
\end{figure}

In Figure~\ref{fig:multimap}, the integrated emission of each observed group of \cch~hyperfine transitions is shown. Each group of transitions follows broadly the same distribution with the N=2-1 group being the brightest. Most strikingly, the majority of the \cch~emission does not appear to follow the full extent of the outflow traced in CO by \citet{Krieger2019}. The N=1-0 emission traces the SW streamer which has been observed in CO \citep{Walter2017Dense253} (see Figure~\ref{fig:1-0map}) but the outflow is otherwise not apparent.\par
The majority of the emission arises from five clumps that are therefore the focus of this work. The position of each clump corresponds to one of the giant molecular clouds (GMCs) identified in \citet{Leroy2015} and so their numbering system is adopted. However, it should be noted that, at higher resolution, GMC 4, 5 and 6 have been resolved into smaller structures and consist of several dense objects that are blended in larger beams \citep{Ando2017DiverseImageS,Mangum2019}. \par
The positions of the clumps were first approximately identified by eye and then five 2D dimensional Gaussian distributions were fit to the N = 2--1 moment 0 map using those positions as initial guesses for the central coordinates. The sizes and central positions of the Gaussians were then simultaneously adjusted to minimize the discrepancy between their sum and the moment 0 map. This allowed the positions and sizes of the five clumps to be fit, taking into account the overlap between them. These five objects are the brightest GMCs in the region when observed in both the continuum and the CO 2-1 transition \citep{Sakamoto2011}.\par
In order to approximate an angular size for the GMCs, the FWHM of a symmetric Gaussian with the same solid angle as the fitted 2D Gaussian was calculated. This was taken to be the convolved size of the source in the beam ($\Omega_{S*B}$). From this, the filling factor can be calculated as
\begin{equation}
    \eta_{ff}=\frac{\Omega_{S*B}-\Omega_{B}}{\Omega_{S*B}},
\end{equation}
which is a simple substitution of $\Omega_{S*B}$ into the more standard $\frac{\Omega_s}{\Omega_{S*B}}$ \citep{Martin2019} where $\Omega_{B}$ is the solid angle of the beam. These filling factors were used to correct the intensity of all spectra extracted from the GMCs in this work. The positions of the GMCs, their convolved angular sizes and the filling factors are given in Table~\ref{table:gmcs}.\par
\begin{table}[]
    \centering
    \caption{Details of the GMCs observed in \cch~and studied in this work including co-ordinates, convolved size and beam filling factor.}
    \begin{tabular}{rllrr}
    \toprule
     {\bf GMC} & {\bf RA / "} & {\bf Dec / "} & {\bf $\Theta_{S*B}$ / "} & {\bf $\eta_{ff}$} \\
    \midrule
       3 &  32.826 &  -21.399 &            2.5 &         0.58 \\
       4 &  33.000 &  -19.783 &            2.5 &         0.58 \\
       5 &  33.183 &  -17.498 &            2.4 &         0.56 \\
       6 &  33.344 &  -15.872 &            2.4 &         0.55 \\
       7 &  33.637 &  -13.446 &            2.1 &         0.42 \\
    \bottomrule
    \end{tabular}
    \tablefoot{Coordinates are given as arcsecond offsets from 00$^h$47$^m$00$^s$,-25$^\circ$ 17' 00".}
    \label{table:gmcs}
\end{table}
Whilst most emission is confined to the GMCs, there is also extended emission over the CMZ. In Figure~\ref{fig:multimap}, it is clear the emission above the 3$\sigma$ level generally extends far past the GMCs, covering the CMZ region traced by other molecules \citep{Sakamoto2011}. The CMZ emission is strongest and has the largest extent in the N = 1--0 emission. All beam sizes are 1.6\arcsec so the wider spatial distribution of the N=1-0 emission is not a beam size effect. Therefore, there must be a gas component present that is sufficiently different in nature to the GMCs to excite N=1-0 emission more strongly than the others. Figure~\ref{fig:1-0map} shows the extent of this emission which is referred to as the extended emission going forward.\par
\subsection{Characterizing the extended emission}
\label{sec:spectralmodeling}
The focus of this work is on the \cch~emission from the GMCs. However, one would expect some contribution from the extended emission to any spectra extracted from the GMC positions because the map in Figure~\ref{fig:1-0map} shows the CMZ surrounds the GMCs. Moreover, preliminary modeling work showed that the N=1-0 and N=4-3 lines could not be simultaneously fit by a single component in any spectrum.\par
\begin{figure*}
    \centering
    \includegraphics[width=\textwidth]{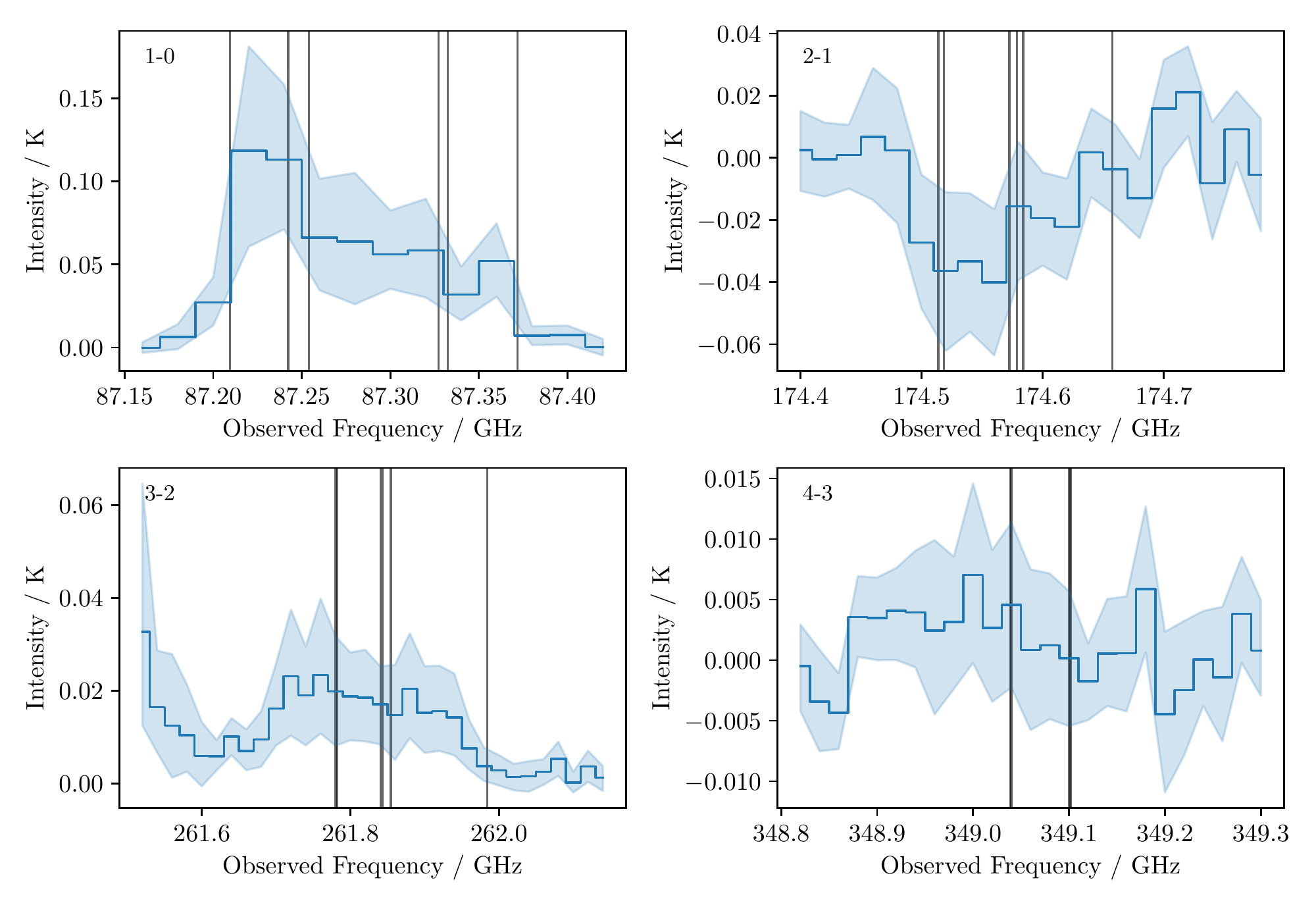}
    \caption{Median stack of all spectra sampled from the extended region of the CMZ. The median is shown as a line and the shaded region gives the 95\% confidence interval on this value. The black squares in Figure~\ref{fig:1-0map} indicate the positions from which spectra were extracted for this composite spectrum. The black lines indicate the positions of the \cch~components in each N transition, Doppler shifted by \SI{255}{\kilo\metre\per\second}, the LSR velocity of \gal.}
    \label{fig:extendedspectra}
\end{figure*}

In order to attempt to reduce the degeneracy inherent in fitting multiple gas components to a spectrum, the extended emission was characterized first to obtain constraints on possible values that could be used in the GMC fit. In Figure~\ref{fig:extendedspectra}, spectra from random positions throughout the extended region are shown. They have been resampled to common frequency bins and median stacked without velocity shifting. The positions from which they were extracted are indicated by black squares in Figure~\ref{fig:1-0map}. \par
These spectra are not directly analysed and are meant only to show the general trend of the extended emission from the CMZ. They are strongest in the N=1-0 transition and become successively weaker at higher N, with the 2--1 line being anomalously weak. This same trend can be seen in the moment 0 maps. Thus the properties of the CMZ gas can be constrained by demanding the emission predicted by radiative transfer model follows this trend.\par%
\begin{table}
    \centering
    \caption{RADEX parameters used to constrain the gas properties of the extended emission component.}
    \begin{tabular}{lcc}
    \toprule
    {\bf Variable}  & {\bf Trialled Range} & {\bf Constrained Range }\\
    \midrule
        Gas Density / \si{\per\centi\metre\cubed} & \num{e2}-\num{e8} & \num{3.5e3}-\num{1.7e4}  \\
        N(\cch) / \si{\per\centi\metre\squared}& \num{e12}-\num{e18} & \num{e13}-\num{e18}\\
        Gas Temperature / K & 10-300  & 50-300 \\
    \bottomrule
    \end{tabular}
    \label{table:extended}
\end{table}
A large grid of RADEX models was run in which the gas density, column density and gas temperature were varied. The trialled ranges of these variables are given in Table~\ref{table:extended}. The parameter space that would be appropriate for the extended emission was then found by utilising two simple constraints. First, all models where the peak N=1-0 emission was less than \SI{0.03}{\kelvin} were rejected because this would be less than three times the RMS noise in the CMZ spectra. Second, any model where the N=1-0 emission was not the strongest of the four detected lines was also rejected as the N=1-0 lines are the brightest part of the extended emission. The parameter values of the remaining models define a parameter space that would be appropriate for the extended emission. \par
\begin{figure}
    \centering
    \includegraphics[width=0.45\textwidth]{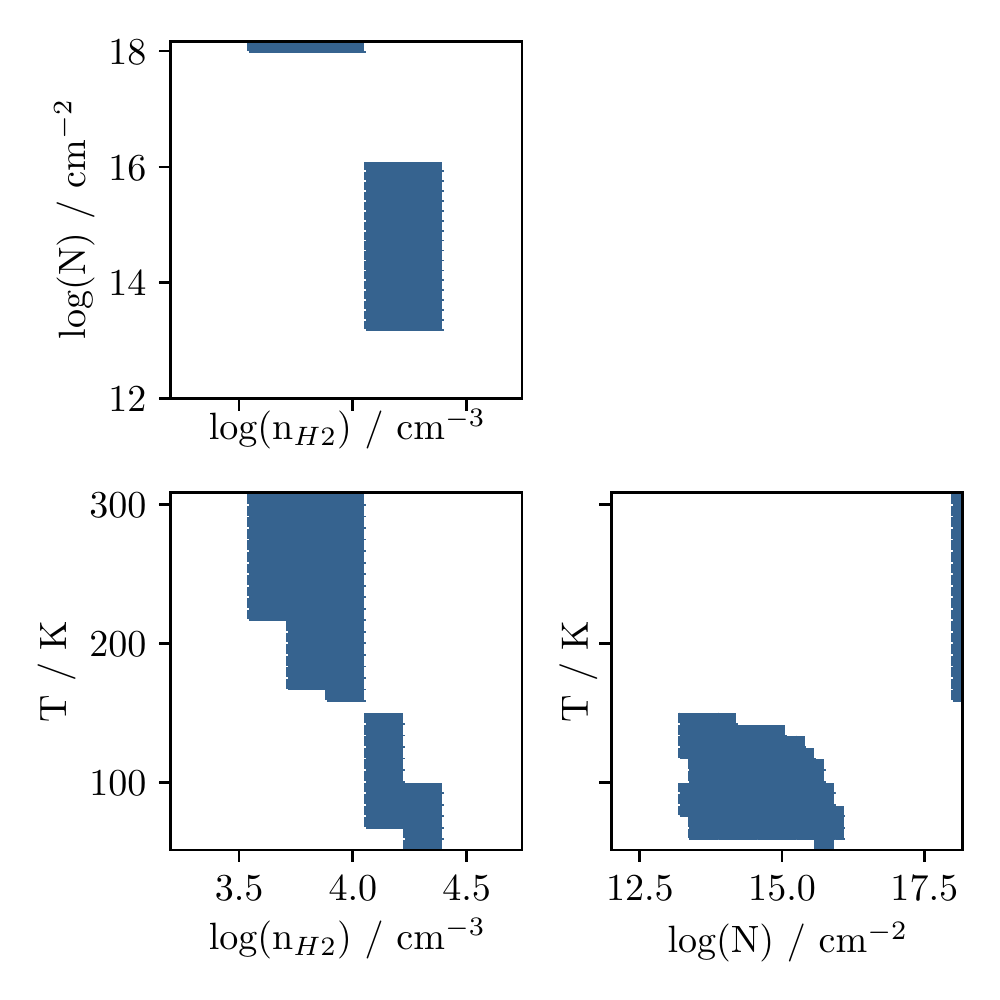}
    \caption{Model parameters that fit constraints from the extended emission of the CMZ. Each point represents the parameters of a model where the N=1-0 line was the strongest line and was above noise.}
    \label{fig:cmz-params}
\end{figure}

Table~\ref{table:extended} gives the parameter ranges that conformed to the above constraints. This is also illustrated in Figure~\ref{fig:cmz-params} where the points indicate a set of parameters that match our constraints. From this figure, relationships between the variables can be seen. Given the weak constraints from the data used for this procedure, the parameters are not well constrained. In particular, the column density is essentially unconstrained as the relative strengths of lines is all that was used rather than the absolute values.\par
However, the constraints on the gas properties are important. The gas must be at a low density between \SI{3.5e3}{\per\centi\metre\cubed} and \SI{1.7e4}{\per\centi\metre\cubed}. It must also be warm as all models that fit our loose constraints have a gas temperature \textgreater 50 K. This is similar to the PDR regions in which \cch~is observed in our own Galaxy \citep{Pilleri2013SpatialR2,Cuadrado2015ThePDR}. Further, these limits can be used to inform the GMC modeling in Section~\ref{sec:gmc-models}, breaking the degeneracy inherent in fitting multiple gas components to a single spectrum.\par
\subsection{Characterizing the GMCs}
\label{sec:gmc-models}
In order to model the GMC spectra, two gas components were used for each GMC. The first was limited to the constrained range of parameters given in Table~\ref{table:extended} to allow for the extended emission contributing to the spectra. The second component simply had to have a larger density than the extended emission and represents the component from the GMC itself. \par
The fits were performed via a Bayesian inference procedure in which emcee \citep{Foreman-Mackey2013} was used to evaluate the posterior distribution of the parameters assuming flat priors and a Gaussian likelihood. The error on each channel was taken to be the spectral noise and an absolute calibration uncertainty of 15\% (see \citealt{Martin2021ALCHEMI} for details) of the channel intensity added in quadrature. Thus the reported most likely values are equivalent to those found through $\chi^2$ minimization but the reported uncertainties are based on the probability distribution of the parameters given the measured spectra.\par
\begin{figure*}
    \centering
    \includegraphics[width=\textwidth]{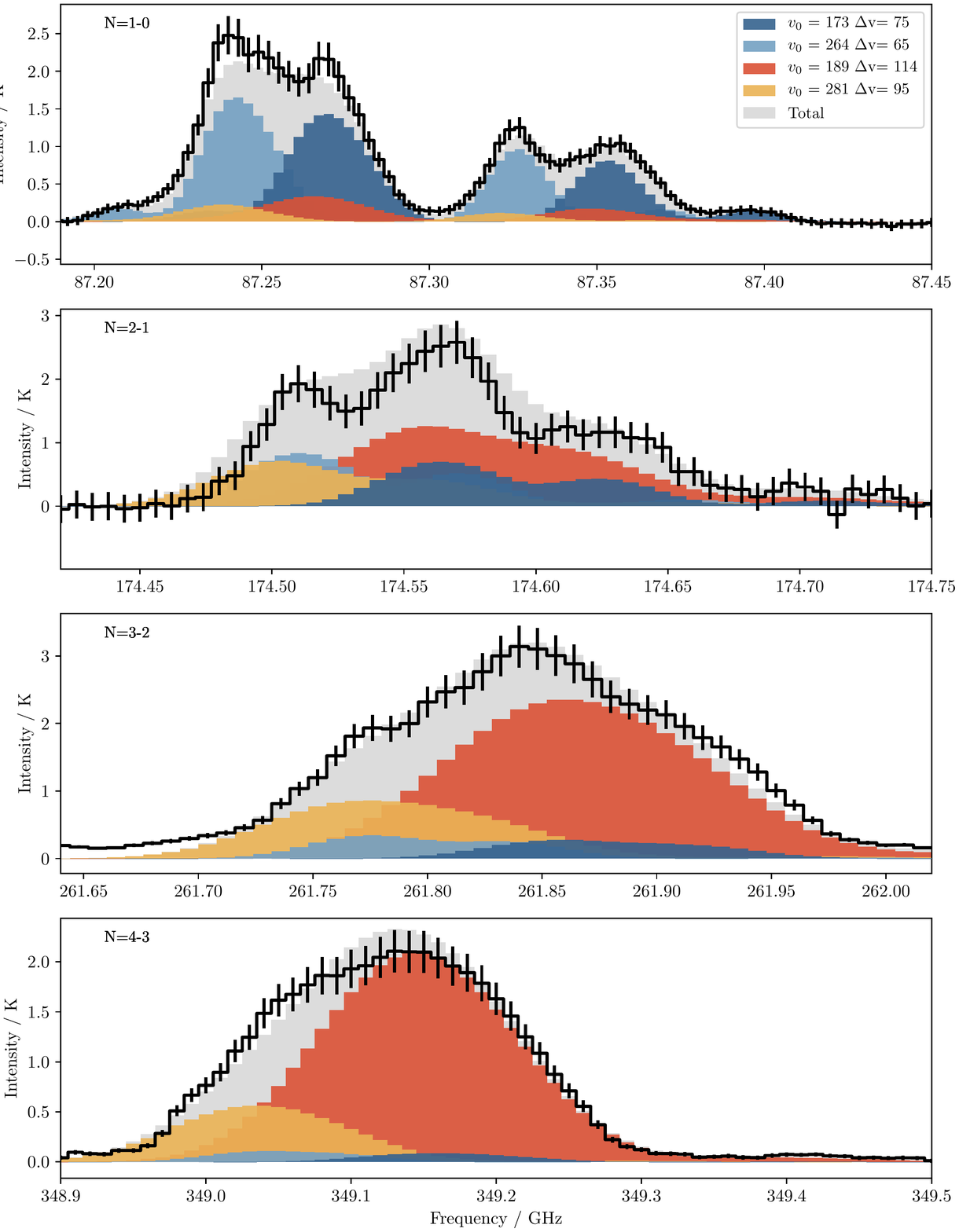}
    \caption{Best fit model for GMC 5. The black trace shows the measured spectra with error bars. The histograms show the spectral model components in colour and their total in grey. The legend gives the central velocity ($V_0$) and linewidth ($\Delta$V) of each component. The two blue components are the low density CMZ contribution and the others are the high density GMC components. }
    \label{fig:gmc5fit}
\end{figure*}

One special case was GMC 5 for which the best fit is shown in Figure~\ref{fig:gmc5fit}. This GMC very clearly has two distinct velocity components, possibly due to the fact that GMC 5 actually comprises at least four clumps with varying gas velocities \citep{Ando2017DiverseImageS}. Despite this known substructure, we chose to fit the two velocity components that are apparent in the data; one with a central velocity of less than \SI{200}{\kilo\metre\per\second} and one with a velocity larger than \SI{200}{\kilo\metre\per\second}. Since the CMZ contaminates both, we further add two low density components. This gives a total of four components: one low density and one high density for each velocity range.\par
The spectra fits in Figure~\ref{fig:gmc5fit} and Appendix~\ref{app:spectra} show that the assumption that the extended emission is contaminating the GMC spectra are justified. The results of the fitting in these figures show the N=1-0 line is dominated by the extended emission component as expect from the fact the extended emission is brightest in this component. \par
The focus of this work is on the GMC emission and so only the values of the high density component are reported in the following sections. The parameters describing the extended components are therefore treated as nuisance parameters and are simply marginalized over. As a result, the reported uncertainty on the GMC component parameters contains the uncertainty from the degeneracy inherent in fitting multiple components.
\subsubsection{\cch~abundance}
\label{sec:columndensity}
The inferred properties of the GMCs, excluding the extended emission, are presented in Table~\ref{table:spectralfits}. For each GMC, the single best fit value of each parameter is presented along with the interval containing 67\% of the probability density of the marginalized posterior distribution of that parameter. This can be thought of as a 1$\sigma$ interval. The RADEX derived column density of \cch~is fairly uniform across the GMCs, each having a best fit value in the range \SIrange[]{e15}{e16}{\per\centi\metre\squared}.\par
\begin{table*}
\centering
\caption{Parameter estimates with $1\sigma$ range from the spectral model fitting to each GMC.}
\begin{tabular}{||l|rl|rl|rl|rl|rl||}
\toprule
GMC  & \multicolumn{2}{l|}{N / 10$^{15}$ cm$^{-2}$} & \multicolumn{2}{l|}{$n_{H2}$ / 10$^5$ cm$^{-3}$} & \multicolumn{2}{l|}{$T_{kin}$ / K} & \multicolumn{2}{l|}{V / km s$^{-1}$} & \multicolumn{2}{l||}{$\Delta$ V / km s$^{-1}$} \\
Component             &                Best Fit &     Range &                    Best Fit &     Range &      Best Fit &       Range &        Best Fit &        Range &                 Best Fit &        Range \\
\midrule
            7 &                     1.5 &   1.1-1.5 &                         6.9 &  1.5-12.1 &          31.7 &   27.7-109.6 &           173.8 &  171.9-175.4 &                     84.1 &   81.6-90.1 \\
            6 &                     3.7 &  3.5-14.1 &                         1.2 &   0.9-4.0 &         - &   \textgreater 70.3.9 &           182.5 &  181.0-185.6 &                     68.4 &   65.9-78.0 \\
    5 - low V &                     4.0 &   1.9-4.1 &                         1.1 &   1.1-2.6 &         - &  \textgreater 109.4.5 &           189.0 &  167.3-194.1 &                    113.8 &  90.2-120.0 \\
   5 - high V &                     1.5 &   1.0-3.7 &                         2.8 &  1.0-18.2 &          - &   \textgreater 29.0 &           280.7 &  228.7-288.4 &                     94.8 &  80.7-140.9 \\
            4 &                     4.2 &  3.9-10.8 &                         1.0 &   0.3-1.5 &         - &  \textgreater 130.3 &           246.3 &  245.1-251.6 &                     80.8 &   76.8-80.9 \\
            3 &                     2.1 &   1.9-2.5 &                         1.1 &   0.9-4.8 &         - &   \textgreater 46.9 &           282.0 &  281.2-282.9 &                     58.9 &   57.4-61.2 \\
\bottomrule
\end{tabular}
\tablefoot{Where the marginalized probability distribution for a parameter becomes approximately flat above a certain value, no best fit value is given and lower limits are provided.}
\label{table:spectralfits}

\end{table*}
However, if the total H$_2$ column density of the cloud is considered, the \cch~abundances are more varied. Using the H$_2$ column density measured from the continuum by \citet{Mangum2019} towards each GMC and assuming the H$_2$ and \cch~trace the same gas, we calculate the \cch~fractional abundances which are given in Table~\ref{table:abundances}. These vary in the range \num{3.6e-10} - \num{1.7e-8}. \par
\begin{table*}
    \centering
\caption{Most likely fractional abundance of \cch and likely bounds}
\begin{tabular}{lcccc}
\toprule
{\bf GMC} &        {\bf X(\cch)} &  {\bf Lower Bound} & {\bf Upper Bound } & {\bf N$_{H2}$ / \SI{e23}{\per\centi\metre\squared}}\\
\midrule
            7 & \num{1.1e-08} & \num{6.6e-09} & \num{1.5e-08} & \num{1.3}\\
            6 & \num{2.5e-09} & \num{2.2e-09} & \num{9.6e-09} & \num{14.7}\\
    5 - low V & \num{1.8e-09} & \num{8.3e-10} & \num{2.1e-09} & \num{21.8}\\
   5 - high V & \num{7.1e-10} & \num{4.3e-10} & \num{1.7e-09} & \num{21.8}\\
            4 & \num{2.8e-09} & \num{2.4e-09} & \num{7.2e-09} & \num{15.0}\\
            3 & \num{2.0e-09} & \num{1.7e-09} & \num{2.5e-09} & \num{10.2}\\
\bottomrule
\end{tabular}
\tablefoot{The lower and upper bound of parameter range covering 67\% of the probability density for this value considering the probability distributions of the \cch~and H$_2$ column densities. The column density values are taken from Table 6 of \citet{Mangum2019}}
\label{table:abundances}
\end{table*}
\subsubsection{Gas properties}
\label{sec:gasprops}
In general the temperature of the gas is poorly constrained by our \cch~data, with the marginalized posteriors for the temperature being approximately flat in the range 50 to 300 K. This is to be expected given that the highest upper state energy of the detected transitions is \SI{41.9}{\kelvin} and thus we should not be sensitive to changes in temperature much above this value. \citet{Mangum2019} also found temperatures \textgreater \SI{50}{\kelvin} on scales similar to the GMCs using measurements of H$_2$CO emission. \par
On the other hand, the gas densities are relatively well constrained and consistent with other measurements in the region. At $\sim$\SI{e5}{\per\centi\metre\cubed}, the gas density is similar to the high end of the range found using dust masses from \citet{Sakamoto2011} and the values measured at higher resolution by \citet{Leroy2018} also from dust.
\section{The origin of \cch~in the \gal~GMCs}
\label{sec:chemicalmodels}
In the previous section, column densities were derived for each GMC, excluding the extended emission. In this section, those column densities and the corresponding fractional abundances are used to constrain chemical models to investigate the origin of \cch~in these clouds.
\subsection{PDR chemistry}
\cch~is ubiquitous in galactic PDRs \citep{Lucas2000ComparativeC_3H_2} and so we explore the possibility that the observed emission arises from the low $A_v$ outer edges of the GMCs. Therefore, a grid of PDR models was run using UCL\_PDR\footnote{\url{https://github.io/UCL_PDR}} \citep{Bell2006,Priestley2017modellingNebula}, a code that has been extensively benchmarked \citep{Rollig2007}. In these models, a 1D cloud of gas in equilibrium was considered with a variety of physical parameters given in Table~\ref{table:pdrgrid} assuming a uniform density. The model iteratively solves the temperature and chemistry considering a variety of heating and cooling processes as well as 215 chemical species interacting through $\sim$2900 reactions until an equilibrium is reached.\par
\begin{table}
    \centering
    \caption{Parameters varied for PDR chemical model.}
    \begin{tabular}{lcc}
    \toprule
    {\bf Variable} & {\bf Symbol} & {\bf Range }\\
    \midrule
        Gas Density & $n_{H2}$ & \num{e4}-\SI{e6}{\per\centi\metre\cubed}  \\
        External Radiation Field &$G$ & 1-\num{e5} Habing\\
        cosmic-ray ionization rate & $\zeta$ & 1, \num{e3} $\zeta_0$\\
    \bottomrule
    \end{tabular}
    \tablefoot{$\zeta_0$ = \SI{1.3e-17}{\per\second}.}
    \label{table:pdrgrid}
\end{table}
The model produces fractional abundances as a function of distance into the GMC or, equivalently, visual extinction. Therefore, the species' column densities can also be calculated by limiting the model to the maximal H$_2$ column density measured towards the region and integrating the \cch~density over that range. \par

\begin{figure*}
    \centering
    \includegraphics[width=\textwidth]{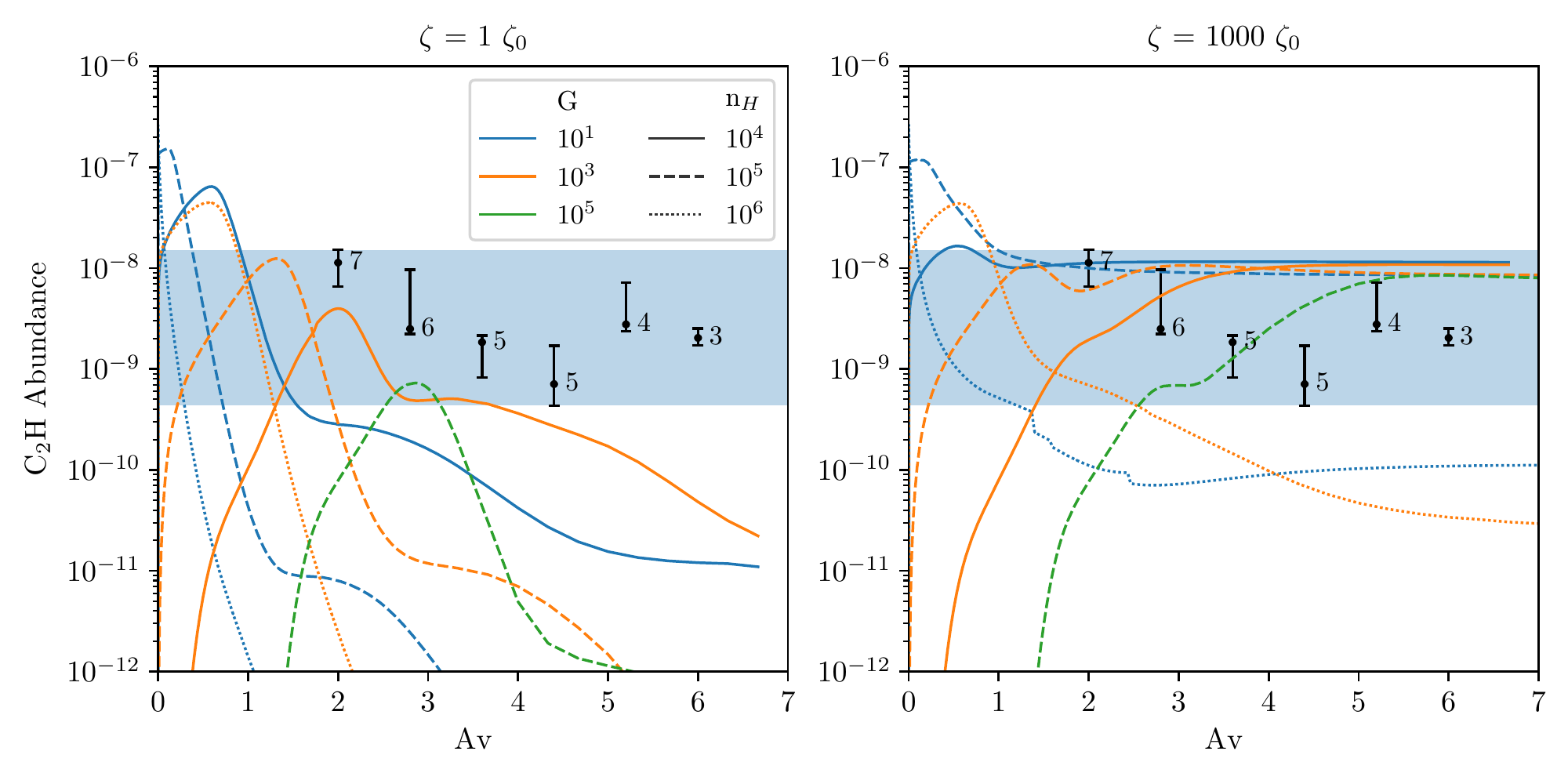}
    \caption{Fractional abundance as a function of depth into the cloud from PDR modeling and observations. The depth into the cloud is given as the equivalent visual extinction. The black dots are numbered according to the GMC they represent and error bars show the observed abundance towards that GMC, their position on the x-axis is arbitrary and does not represent the cloud sizes. The lines show UCL\_PDR models with different parameters. The shaded band covers the abundance range of all GMC measurements. In the low cosmic-ray ionization rate models (left), the abundance decreases with distance into the cloud so teh total column density is less than the observed column density despite the high abundances reached at low $A_v$.}
    \label{fig:pdrgrid}
\end{figure*}

The fractional abundance of \cch~as a function of depth in these models is given in Figure~\ref{fig:pdrgrid}. The key result is that whilst many models reach \cch~abundances compatible with those measured in \gal~at low $A_v$, none are capable of explaining the observed column density of $\sim$\SI{e15}{\per\centi\metre\squared} if a uniform medium is assumed. In fact, assuming a maximum H$_2$ column density of \SI{2.2e24}{\per\centi\metre\squared} (the largest measured by \citealt{Mangum2019} towards one of the GMCs discussed here), the largest possible \cch~column density from a model with the canonical Milky Way cosmic-ray ionization rate of $\zeta_0$=\SI{1.3e-17}{\per\second} is \SI{6e13}{\per\centi\metre\squared}. This is over an order of magnitude too low to match any GMC.\par
This is because the abundance reduces with depth into the cloud so only the outer edges contribute to the column. One solution to this would be to invoke clumpiness. However, for PDR chemistry to be responsible for the observed \cch~emission, the average A$_v$ would need to be less than 2 magnitudes despite the fact the column density in these clouds would give an A$_V$ \textgreater 500 magnitudes.\par
If, on the other hand, an enhanced cosmic-ray ionization rate is introduced, the model abundances of \cch~become enhanced and constant with respect to $A_v$. This enhancement is due to the high ionization rate creating an environment where the ionization fraction is high, much like a PDR, even in high $A_v$ regions of the model. In the models, \cch~is produced primarily through the chain,
\begin{enumerate}
    \item C$^+$ + CH$_2$ $\longrightarrow$ C$_2$H$^+$ + H
    \item H$_2$ + C$_2$H$^+$ $\longrightarrow$ C$_2$H$_2^+$ + H
    \item C$_2$H$_2^+$ + e$^-$ $\longrightarrow$ \cch~+ H
\end{enumerate}
regardless of whether C$^+$ is produced through photochemistry or cosmic ray ionization. For a rate of 1000 $\zeta_0$, many models give a \cch~abundance of $\sim$\num{e-8} regardless of depth into the cloud. This produces column densities in line with those observed. Therefore, ruling out the possibility that the the GMCs are so clumpy that the average visual extinction is less than 2 magnitudes, it can be concluded that the \cch~emission observed in \gal~does not primarily originate from the PDR regions in the clouds. If that is the case, the \cch~emission must come from within the clouds where some process maintains the \cch~abundance at high visual extinctions and this is explored in the following sections.\par
\subsection{Dense cloud chemistry}
\label{sec:dark-models}
Given the likelihood that \cch~emission observed towards the GMCs in \gal~comes from within those clouds rather than the UV irradiated skin, it becomes appropriate to model the chemistry with a dense cloud model where photo-processes are assumed to be negligible. UCLCHEM\footnote{\url{https://uclchem.github.io}}, a gas-grain chemical code \citep{Holdship2017} was used for this purpose.\par
It is assumed that since the extinction of UV photons is the only depth dependent process, the entire interior of a GMC can be modeled as a single point with an $A_v$ that is sufficiently high to make all UV processes negligible. Therefore, a grid of single point models was run with an $A_v$ of 10 mag which is sufficient to reduce the local UV field in the model to zero. This grid covered a range of gas densities, temperatures, and cosmic-ray ionization rates to represent different GMC conditions which are given in Table~\ref{table:uclchemgrid}.\par
Each GMC model used initial abundances that were generated by a model that started from purely atomic gas with solar elemental abundances \citep{Asplund2009} with silicon depleted to 1\% of its solar value. The initial density was $n_{H_2}$=\SI{e2}{\per\centi\metre\cubed} and the density then increased according to a freefall collapse model to the required density of the GMC model. This provided realistic initial abundances for a GMC that has formed from diffuse gas without assuming values which the chemical network may not be able to produce. Each model was run for 5 Myr so that the abundance over time and the steady-state value could be analysed.\par
\begin{figure*}
    \centering
    \includegraphics[width=\textwidth]{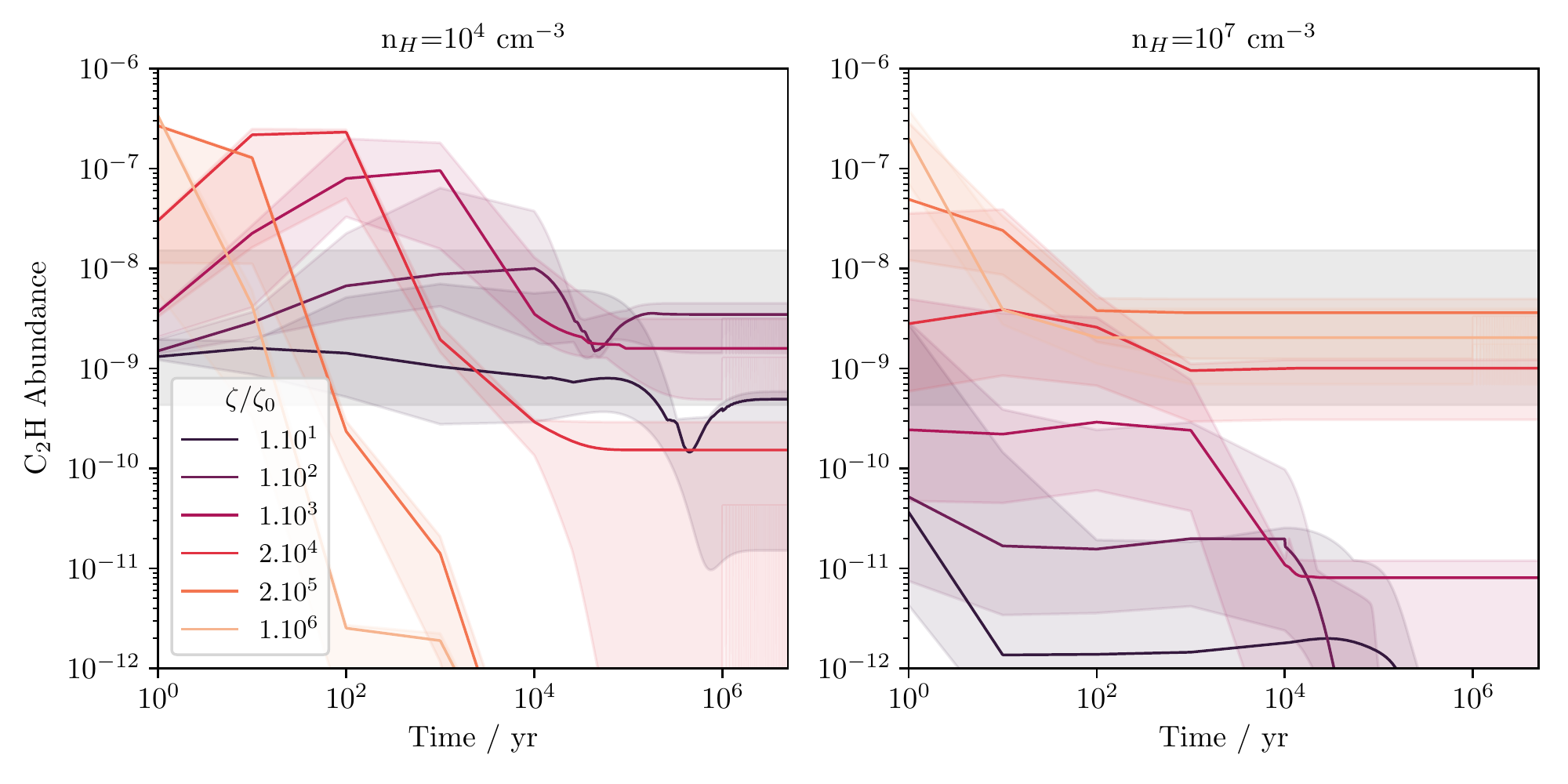}
    \caption{Fractional abundance of \cch~as a function of time in the UCLCHEM models. Each line shows the median value of the abundance for a given cosmic-ray ionization rate as temperature is varied and the shaded regions show the minimum and maximum abundances over that temperature variation. The cosmic-ray ionization rate is given as a multiple of $\zeta_0$=\SI{1.3e-17}{\per\second} and a sample of the lines are given in the legend. The grey shaded area shows the observed range.}
    \label{fig:abundances}
\end{figure*}

\begin{table}
    \centering
    \caption{Parameters varied for dark cloud chemical model.}
    \begin{tabular}{lcc}
    \toprule
    {\bf Variable} & {\bf Symbol} & {\bf Range }\\
    \midrule
        Gas Density & $n_{H2}$ & \num{e4}-\SI{e6}{\per\centi\metre\cubed}  \\
        Gas Temperature & $T_g$ & 50-300 K\\
        cosmic-ray ionization rate & $\zeta$ & 10-\num{e6} $\zeta_0$\\
    \bottomrule
    \end{tabular}
    \tablefoot{$\zeta_0$ = \SI{1.3e-17}{\per\second}.}
    \label{table:uclchemgrid}
\end{table}
The model results in Figure~\ref{fig:abundances} show that equilibrium values are quickly reached. Note that at t=0 yr, the model is already at the post-collapse density. In fact, in many cases, the \cch~abundance becomes constant after as few as \num{e4} years. Further, within the limits of the grid, the steady state abundance of \cch~is typically not strongly affected by the gas temperature. In fact, as long as the abundance of \cch~is sufficiently high to match the observations, the temperature only creates a small variance as shown by the shaded region around each line in the plot. Instead, the combination of the gas density and cosmic-ray ionization rate sets the abundance. \par
\begin{figure}
    \centering
    \includegraphics[width=0.5\textwidth]{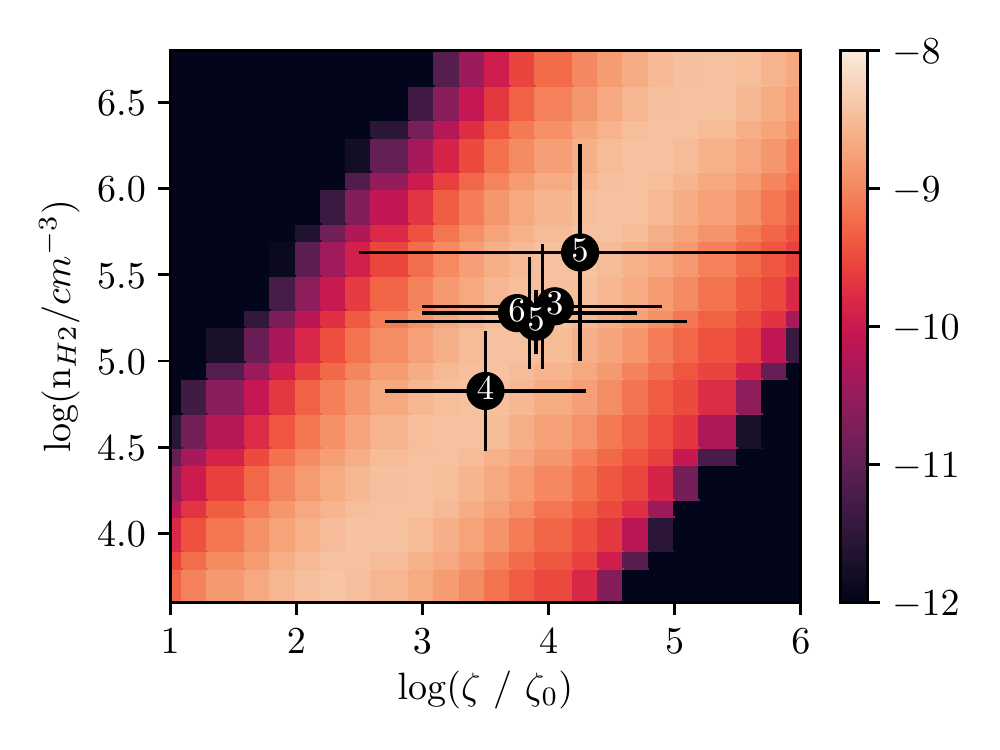}
    \caption{Steady state \cch~abundance shown in colour scale as a function of gas density and cosmic-ray ionization rate in units of $\zeta_0$=\SI{1.3e-17}{\per\second}. The values given are a median over different gas temperatures. Each point represents a GMC with the vertical lines showing the upper and lower limits on the density for that cloud. The horizontal lines span the cosmic-ray ionization rate values that give abundances within the measured range for those density values. GMC 7 does not appear on this diagram as no model produces a large enough abundance of \cch~to match even its observed lower-limit.}
    \label{fig:zetanhgrid}
\end{figure}

If it is assumed the abundances reach steady state, the cosmic-ray ionization rate required to obtain the measured \cch~abundances appears to have a log-log relationship to the density. This is to be expected given that higher densities tend to decrease the total ionization fraction in the models and so a higher cosmic-ray ionization rate is needed to maintain the PDR-like chemistry that produces \cch. Thus, using the estimates of the gas density from Section~\ref{sec:gasprops}, it may be possible to constrain the ionization rate. Figure~\ref{fig:zetanhgrid} shows the steady state abundance of \cch~as a function of the gas density and cosmic-ray ionization rate. Overplotted are points representing each GMC, showing the range of densities found in Section~\ref{sec:gasprops} and the cosmic-ray ionization rate range that gives model abundances within the measured limits for that GMC.\par
GMC 7 presents a problem as no model produces a large enough abundance of \cch~to match even the lower limit for that GMC and hence it is missing from Figure~\ref{fig:zetanhgrid}. However, there is a lot of uncertainty in both the derived fractional abundance and the chemical model. The H$_2$ column density used to convert to abundance was derived using an assumed dust temperature and dust to gas ratio as well as assuming optically thin dust emission. If the resulting column density was underestimated by even a factor of 2, the abundance can be fit. If we further consider uncertainties such as the initial elemental abundance of carbon in the chemical model, a discrepancy of this magnitude is unsurprising.\par
The poor fit of GMC 7 and the fact that the cosmic-ray ionization rate can only be constrained to within a few orders of magnitude for the other GMCs indicates that this analysis should only be considered qualitatively. The fits show that the cosmic ray ionization rate is likely to be high in these regions but the specific values are very uncertain. Each GMC has a lower limit on the cosmic-ray ionization rate of 1500$\zeta_0$ except for the high velocity component of GMC 5 where the density is poorly constrained and so $\zeta$ can be as low as 100$\zeta_0$. Overall, we can conclude that the observed \cch~emission could arise from the chemistry of a cosmic ray dominated region where the cosmic-ray ionization rate must be much higher than standard but cannot constrain it to within an order of magnitude.\par
\subsection{Shock chemistry}
Whilst the previous section indicates cosmic rays may be the main driver of the chemistry producing \cch, it should be noted that other possibilities exist. Most importantly, the model effectively treats cosmic rays as ionization events that are not reduced by the gas column density and so other sources of ionization such as the strong X-ray irradiation in the region \citep{Strickland2002Galaxies} could be at work instead. It is also possible that physical processes such as shocks not included in the previous models could cause an enhanced abundance of \cch.\par
The dense cloud modeling shows that simply heating the gas or increasing the density cannot produce enough \cch. This is illustrated in Figure~\ref{fig:abundances} where the shaded regions show little variation in abundances as the temperature varies between 50 and 300 K and in Figure~\ref{fig:zetanhgrid} which shows the \cch~abundance actually decreases with increasing density for a given cosmic ray ionization rate. However, more complex physical processes could be at work. A natural possibility is turbulence which would affect the chemistry through shocks as low velocity shocks have previously been found to dominate the heating in the region \citep{Martin2006}. These shocks would tend to heat and compress the gas as well as remove material from the grains. Furthermore, in NGC~1068 it was found one possible cause of the high \cch~abundances was in fact shocks \citep{Aladro2013AEnvironment,Garcia-Burillo2017}.\par
To explore this, we run a simple grid of shock models using UCLCHEM's shock module based on the C-shock parametrization of \citet{jimenez2008}. We vary the shock velocity from 5 to \SI{40}{\kilo\metre\per\second}, to trial a range of velocities over which the shock treatment is applicable. We apply these shocks to gas with pre-shock densities between \num{e4} and \SI{e6}{\per\centi\metre\cubed} with initial abundances taken from a collapse model as described in Section~\ref{sec:dark-models}. We vary the pre-shock gas temperature from 10 to \SI{50}{\kelvin} and cosmic-ray ionization rate between 1 and 10 $\zeta_0$ to separate the effects of shocks and high ionization rates. These parameter ranges are summarized in Table~\ref{table:shockgrid}.\par
\begin{table}
    \centering
    \caption{Parameters varied for shock chemical models.}
    \begin{tabular}{lcc}
    \toprule
    {\bf Variable} & {\bf Symbol} & {\bf Range }\\
    \midrule
        Gas Density & $n_{H2}$ & \num{e4}-\SI{e6}{\per\centi\metre\cubed}  \\
        Gas Temperature & $T_g$ & 10-50 K\\
        cosmic-ray ionization rate & $\zeta$ & 1-10 $\zeta_0$\\
        Shock Velocity & $V_s$ & 5-40 \si{\kilo\metre\per\second}\\
    \bottomrule
    \end{tabular}
    \tablefoot{$\zeta_0$ = \SI{1.3e-17}{\per\second}}
    \label{table:shockgrid}
\end{table}
In almost every model, the pre-shock \cch~abundance was an order of magnitude too low to match observations and the passage of a shock permanently reduced the \cch~abundance, meaning most shocks could not reproduce observations. However, a small subset of shock models with a density of \SI{e4}{\per\centi\metre\cubed} and a shock velocity greater than \SI{20}{\kilo\metre\per\second}, had \cch~abundances that were enhanced briefly by the shock passage. In each of these models, there is a short period of $\sim$\num{e5} years where the \cch~abundance is enhanced to the levels observed in \gal. \par
\begin{figure}
    \centering
    \includegraphics[width=0.5\textwidth]{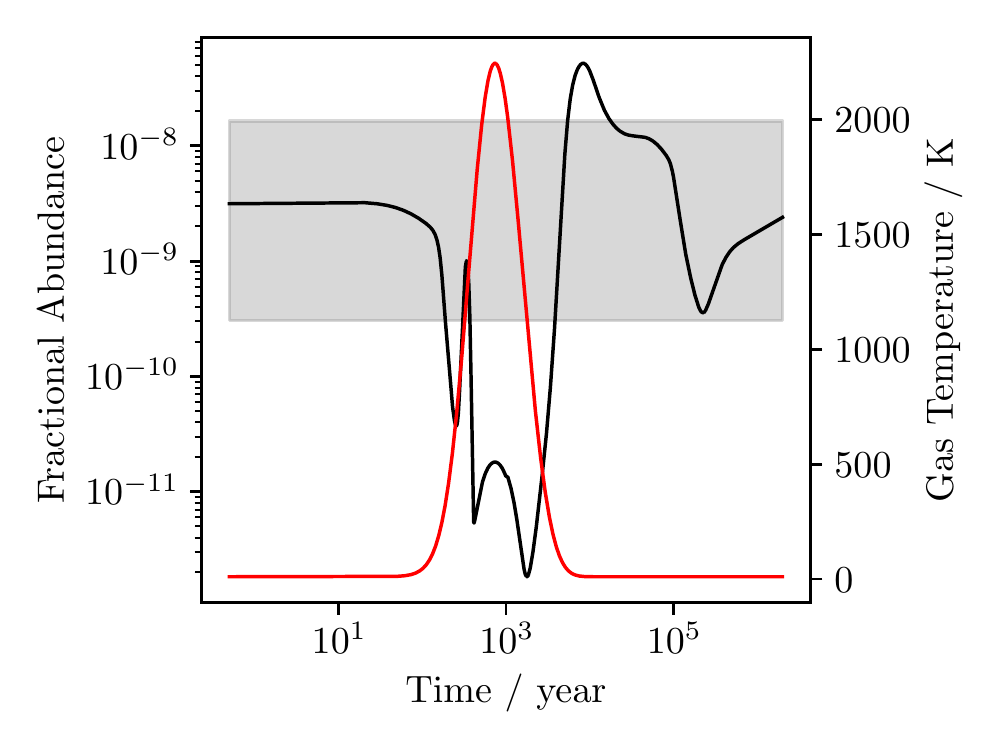}
    \caption{Fractional abundance of CCH through time for a 20 km s$^{-1}$ shock propagating through gas with a pre-shock density of \SI{e4}{\per\centi\metre\cubed} plotted in black, temperature through the shock plotted in red. The range of observed \cch~abundances is shown in grey.}
    \label{fig:shocktrace}
\end{figure}
Naturally, a single simple shock model is difficult to apply to a large cloud that is possibly experiencing multiple small shocks due to turbulent internal motion. However, taking an average over the shock structure, the model abundances of this subset of shock models are all within the lower and upper bounds on the \cch~abundance for at least one GMC as given in Table~\ref{fig:abundances}. Therefore, if one argues that shocks are so ubiquitous in these GMCs that almost all of the gas is constantly in a state of having just been shocked (within ~\num{e5} years) then the overall average \cch~abundance will be within our observational constraints.\par
However, there are several aspects of the GMCs that indicate the required shock conditions are not met in these objects. Our models indicate shocks with a velocity greater than \SI{20}{\kilo\metre\per\second} are necessary but previous work has suggested shocks (\textless \SI{20}{\kilo\metre\per\second}) are common in the region \citep{Garcia-Burillo2000Large-scaleDisk,Martin2006}. Moreover, the post shock density in these models is only \SI{5e4}{\per\centi\metre\cubed} and the gas densities in the GMCs are larger. This is vital as models with higher pre-shock gas densities do not reproduce the observed abundances. Finally, it is unlikely that all of the gas in the GMCs is shocked repeatedly on a timescale of ~\num{e5} years but averaging over longer times in the shock model produces abundances that are lower than those observed. \par
To summarize, a subset of shock models could potentially reproduce the abundances of \cch~observed in these clouds and therefore cannot be ruled out. However, the conditions required do not match those measured for the GMCs and the gas would have to be almost constantly undergoing a shock. Therefore, we consider it unlikely in comparison to the wide range of cosmic ray dominated models that fit the measured abundances.
\subsection{Relationship to other sources}
In the Milky Way, it is typical to observe \cch~with abundances similar to those measured for \gal~in PDRs. However, it is clear from the modeling done in this work that it cannot be the case that \cch~in \gal~arises entirely from PDRs. The observed column densities are far too high and another process must be at play. \par
In this work, we have proposed that a large degree of ionization, likely from cosmic rays, is responsible for the \cch~abundance or, less likely, ubiquitous shocks in the GMC. This is similar to the molecular cloud G+0.693-0.027  which has a \cch~abundance of \num{4e-8}\citep{Bizzocchi2020PropargylimineISM}. The cosmic ray ionization rate is expected to be several orders of magnitude higher than the standard Milky Way value at the Galactic centre. However, the chemistry in the cloud is also thought to be shock dominated \citep{Requena-Torres2006OrganicCores}.\par
A comparison to NGC~1068 is also of interest. In that galaxy, the \cch~emission was observed towards both the starburst ring and along the AGN driven outflow. However, only the N=1-0 transition was observed by \citet{Garcia-Burillo2017} and so the column density could not be well constrained. Assuming temperatures around \SI{50}{\kelvin} and LTE, a fractional abundance of \cch~similar to that measured here for \gal~can be recovered for the starburst ring. However, a true comparison of the star forming regions in these galaxies would require higher N transitions of \cch~to be observed toward NGC~1068 as the majority of emission from the GMCs in \gal~was found in those lines. \par
\cch~was also detected in NGC~1068 at higher abundances in the range \num{e-7} to \num{e-6} around the AGN driven outflow. A variety of physical factors including high UV fields, high cosmic-ray ionization rates and shocks could all produce this abundance for a very short time according to chemical modeling work in \citet{Garcia-Burillo2017}. As a result, the authors conclude the high abundances are maintained by a dynamic environment around the outflow in which gas is constantly resupplied to the interface between the outflow and its surroundings. As a result, a pseudo-steady state is reached where the gas is well fit by the early stages of their chemical models. Whilst this picture makes sense for an outflow interface, it does not apply to the GMCs observed in \gal~which are much more static. \par
Thus, we are tracing fundamentally different gas in each galaxy with \cch. This is evidenced by the fact the majority of the \cch~emission does not appear to follow the outflow of \gal~ as traced in CO by \citet{Krieger2019}. In fact, only the N=1-0 line traces the base of the outflow as found in previous work \citep{Meier2015ALMA253} unlike in NGC 1068 where is traces a large extend of the outflow \citep{Garcia-Burillo2017}. This lack of outflow emission indicates that the energetic processing producing the high abundance of \cch~in NGC~1068's AGN driven outflow is not present in \gal's starburst driven outflow. \gal~appears to represent a third class of \cch~rich gas in which an ionizing process such as cosmic ray ionization maintains an ion dominated chemistry and therefore enhanced abundances of \cch~well into dark regions. It is clearly distinguishable from the outflow case by its lower \cch~abundance and coincidence with dense gas but differentiated from PDR emission by its high column density.
\section{Conclusions}

The emission of the N=1-0 to N=4-3 rotational transitions of \cch~in \gal~was imaged with ALMA at a resolution of 28 pc (1.6\arcsec). Most of the \cch~emission traces the dense gas of the brightest GMCs that have been previously studied in the CMZ of \gal. However, the emission from the N=1-0 transition also traces diffuse gas in the CMZ itself. \cch~does not appear to follow the galaxy's starburst driven outflow.\par
Spectral modeling was used to infer the gas properties of the GMCs. The temperatures were found to be \textgreater \SI{50}{\kelvin} and the gas densities varied from \SIrange[]{e5}{e6}{\per\centi\metre\cubed}. These values are in line with those previously measured in the GMCs. The column density of \cch~was also constrained to be in the range \num{e15}-\SI{e16}{\per\centi\metre\squared} which corresponds to fractional abundances $\sim$\num{e-9}.\par
Chemical modeling showed that despite being enhanced in PDRs, the \cch~emission in these GMCs could not entirely arise from the photon dominated outer regions of these dense clouds but must instead come from within the clouds. \cch~in these GMCs most likely arises from gas where the ionization fraction is kept high by some ionizing process such as cosmic rays. Alternatively, ubiquitous shocks could be responsible for the measured \cch~abundance but this would require the entire gas to be shocked on such a short timescale that this is unlikely.\par
If a high cosmic-ray ionization rate is responsible for the \cch~abundance, the ionization rate in each GMC is constrained to within a few orders of magnitude only. The values for each GMC vary from \num{e3} to \num{e6} $\zeta_0$. A more sensitive probe of the cosmic-ray ionization rate should be utilized in these regions to confirm the presence of a high ionization rate and better constrain its value.

\begin{acknowledgements}
      We thank the anonymous referee for their report which led to the improvement of this manuscript. JH and SV are funded by the European Research Council (ERC) Advanced Grant MOPPEX 833460.vii VMR and LC are funded by the Comunidad de Madrid through the Atracci\'on de Talento Investigador (Doctores con experiencia) Grant (COOL: Cosmic Origins Of Life; 2019-T1/TIC-15379). This paper makes use of the following ALMA data: ADS/JAO.ALMA\#2017.1.00161.L and ADS/JAO.ALMA\#2018.1.00162.S. ALMA is a partnership of ESO (representing its member states), NSF (USA) and NINS (Japan), together with NRC (Canada), MOST and ASIAA (Taiwan), and KASI (Republic of Korea), in cooperation with the Republic of Chile. The Joint ALMA Observatory is operated by ESO, AUI/NRAO and NAOJ.
\end{acknowledgements}

%
\bibliographystyle{aa} 
\bibliography{references} %
%

%
\begin{appendix} 
\section{SpectralRadex}
\label{app:spectralradex}
\subsection{Package description}
SpectralRadex\footnote{\url{https://spectralradex.readthedocs.io}} is a Python module that was created in the process of this work and is now generally available through github\footnote{\url{https://github.com/uclchem/SpectralRadex}} and Pypi. As such it is described here in detail.\par
The module comprises two parts: a wrapper for RADEX \citep{VanderTak2007} and a spectral modeling library. Whilst many packages exist for the former purpose, most either compile the original RADEX source code including Fortran 77 COMMON blocks using numpy's F2PY \cite{} or create an interface for calling the compiled RADEX binary. The former has less I/O overhead and should be preferred but the use of COMMON blocks means no copies of the program can be run simultaneously and is therefore problematic for multiprocessing. SpectralRadex solves this issue by rewriting the RADEX source code, updating it to modern Fortran standards and dropping common blocks in favour of Fortran modules before compiling with F2PY.\par
For the latter purpose, SpectralRadex makes use of RADEX to produce non-LTE spectra from RADEX inputs, a set of frequencies at which to evaluate the intensities and the velocity by which all lines should be shifted. The formalism for this is detailed below and borrows heavily from CASSIS \citep{Vastel2015}.
\subsection{Spectral modeling Formalism}
In order to calculate the emission from a single molecular transition as a function of frequency, the excitation temperature and the optical depth are required. The brightness temperature is given by,

\begin{equation}
	T_B = [J_{\nu}(T_{ex})-J_{\nu}(T_{BG})](1-\exp(-\tau_v))
\end{equation}

where $J_v$ is the radiation temperature, $T_{ex}$ is the excitation temperature, $T_{BG}$ is the background temperature, and $\tau_v$ is the optical depth as a function of velocity. The radiation temperature is simply,
\begin{equation}
	J_{\nu}(T_{ex})= \frac{\frac{h\nu}{k}}{\exp{\frac{h\nu}{kT_{ex}}}-1}
\end{equation}
where $\nu$ is the frequency, $h$ is Planck's constant and $k$ is Boltzmann's constant. The optical depth, can then be calculated by converting frequency to equivalent velocity shift using the rest frequency of the transition and then assuming a Gaussian line profile

\begin{equation}
    \tau_v = \tau_0 e^{\left(-4ln(2)\frac{(v-v_0)^2}{\Delta v^2}\right)}
\end{equation}
where $\tau_0$ is the optical depth at line centre, $v_0$ is the velocity shift of the emission, and $\Delta v$ is the FWHM of the line. Thus all that is required is to calculate $T_{ex}$ and $\tau_0$ for a given transition and set of physical parameters.\par
This can be achieved through the use of RADEX. For a given set of physical parameters RADEX will provide the optical depth at line centre for every transition and the excitation temperature that gives the correct brightness temperature at line centre. Thus, a non-LTE spectrum can be generated by taking these quantities from an appropriate RADEX model. In the high density limit, this tends to the LTE solution but at lower densities it can deviate significantly.\par
In SpectralRadex, $T_B$ is calculated as a function of frequency for each line and then combined to give the overall spectrum of the molecule. Where lines overlap, SpectralRadex follows \citet{Hsieh2015PropertiesObjects} and uses an opacity weighted radiation temperature:

\begin{equation}
	T_B = \left(\frac{\Sigma_i J{\nu}(T^i_{ex})\tau^i_v}{\Sigma_i \tau^i_v}-J_{\nu}(T_{BG})\right)(1-\exp(-\tau_v))
\end{equation}

\section{The other fits}
\begin{figure*}
    \centering
    \includegraphics[width=\textwidth]{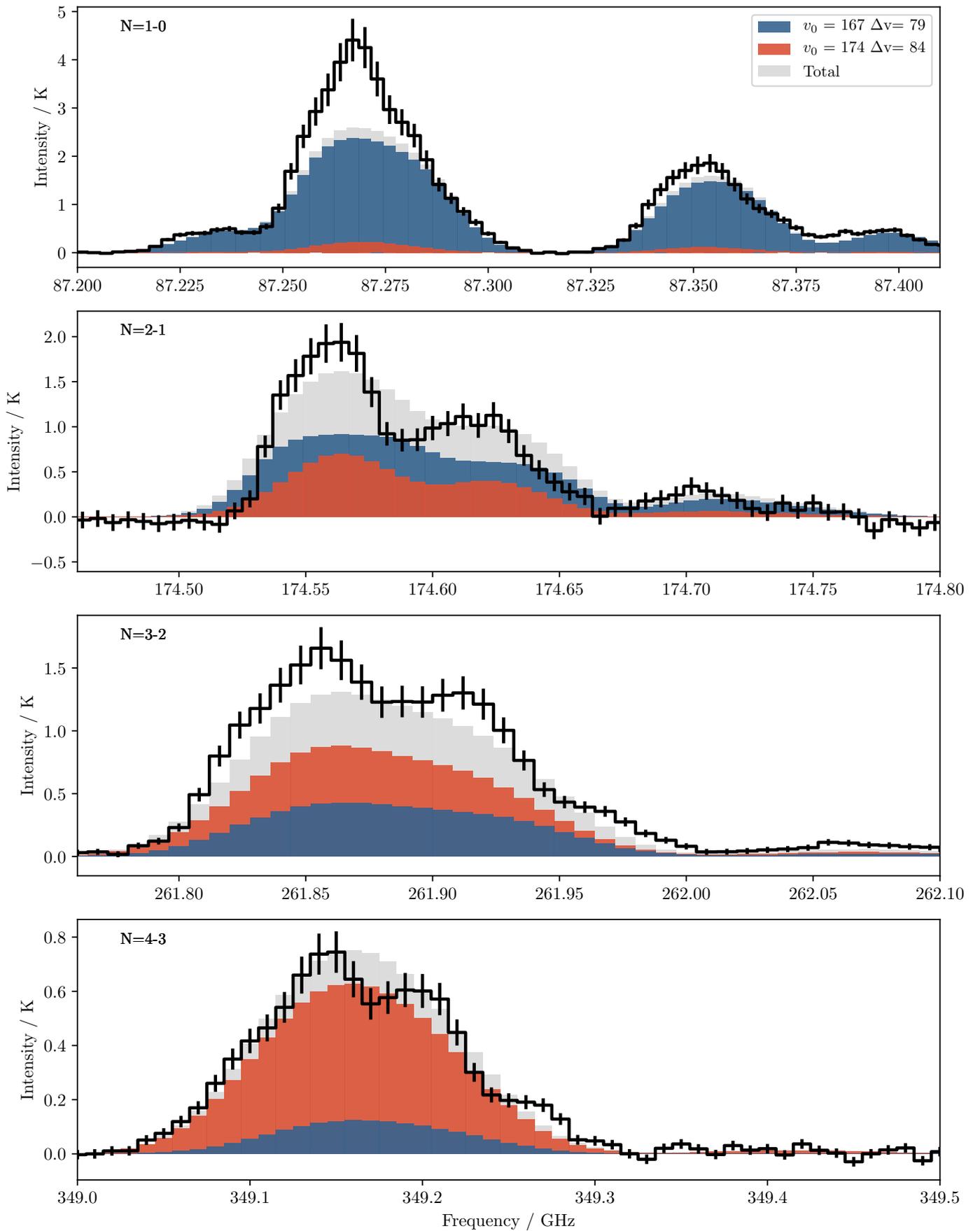}
    \caption{Measured spectrum from GMC 7 in black with error bars. Model spectra generated from most likely values for GMC 7 as filled histograms. The blue component shows the low density extended emission, the red shows the high density GMC emission, and the grey shows the complete model.}
    \label{fig:gmc7}
\end{figure*}

\begin{figure*}
    \centering
    \includegraphics[width=\textwidth]{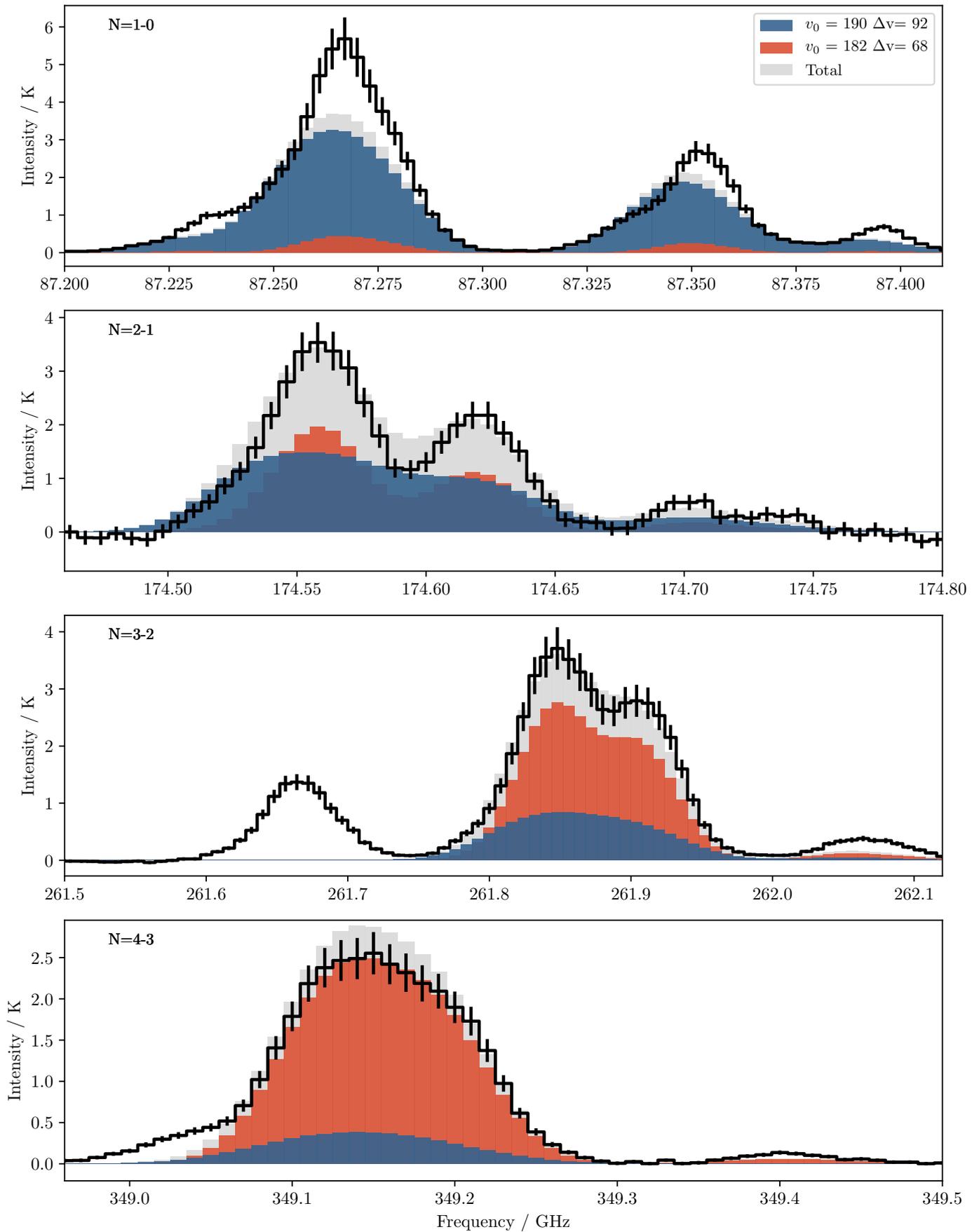}
    \caption{Similar to Figure~\ref{fig:gmc7} for GMC 6}
    \label{fig:gmc6}
\end{figure*}

\begin{figure*}
    \centering
    \includegraphics[width=\textwidth]{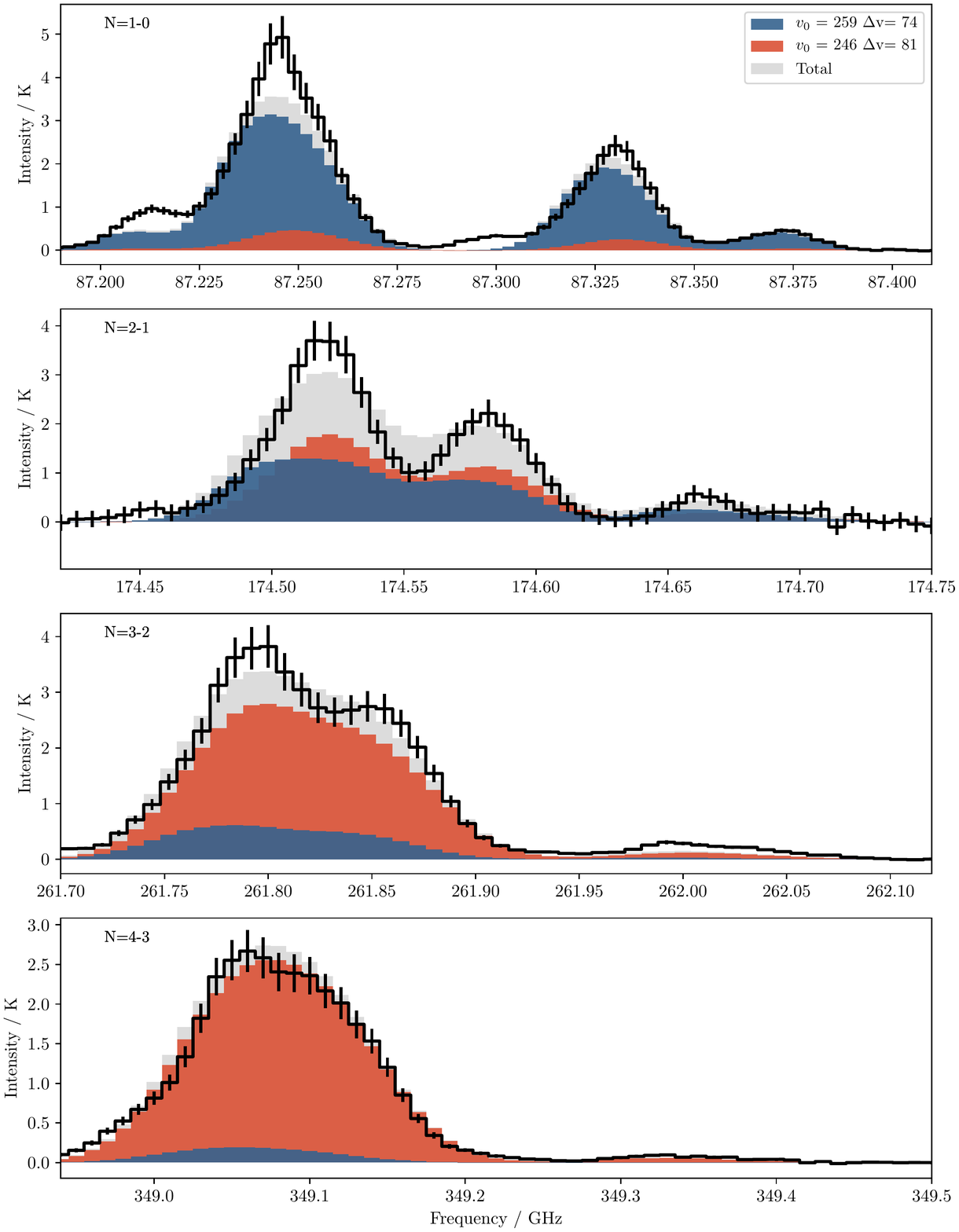}
    \caption{Similar to Figure~\ref{fig:gmc7} for GMC 4}
    \label{fig:gmc4}
\end{figure*}

In this section, the best fit models for each of the GMC spectra are shown except for GMC 5 which is shown in Figure~\ref{fig:gmc5fit}. Figure~\ref{fig:gmc7} shows GMC7, Figure~\ref{fig:gmc6} shows GMC 6 and Figure~\ref{fig:gmc4} shows the best fit to GMC 4.
\label{app:spectra}

\begin{figure*}
    \centering
    \includegraphics[width=\textwidth]{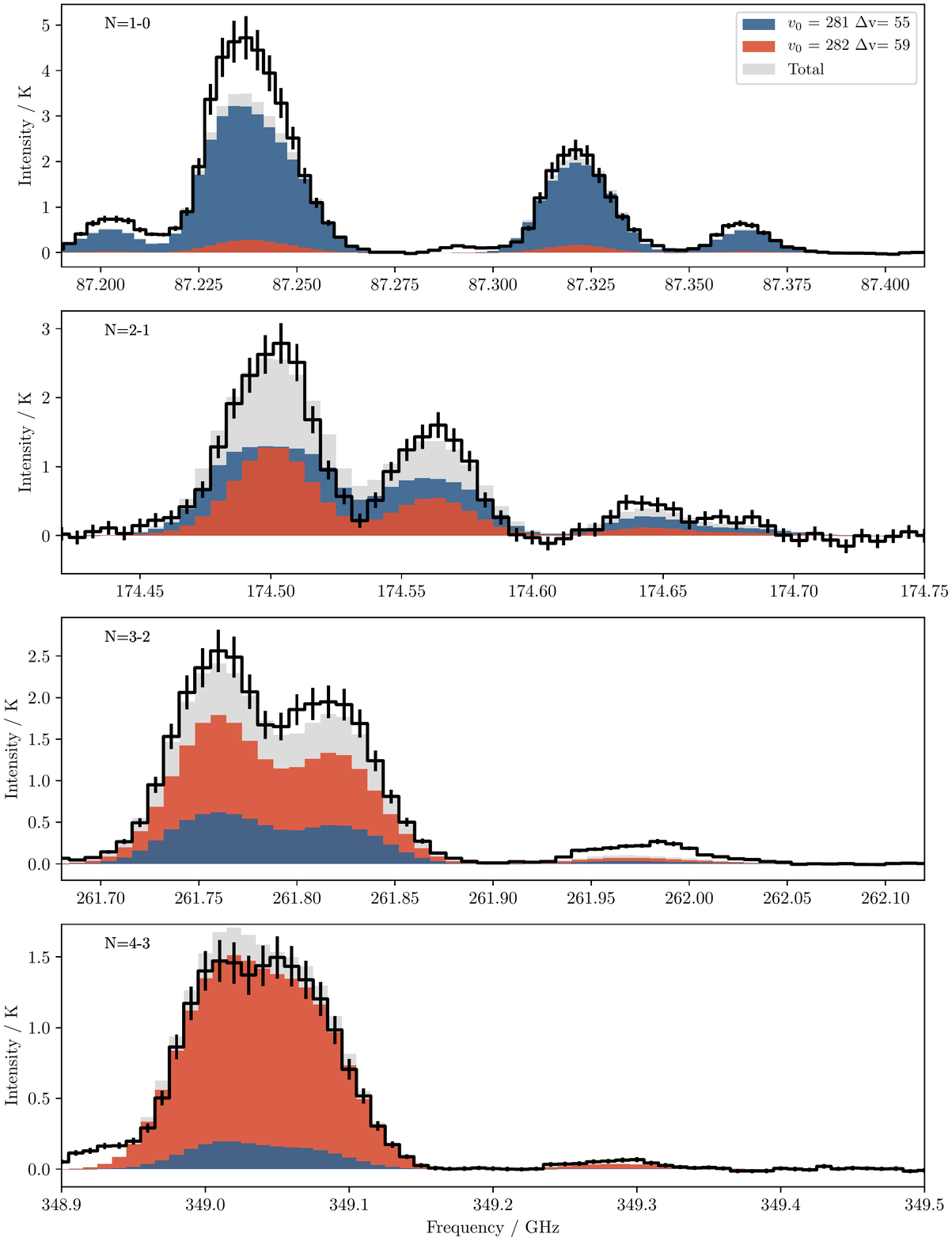}
    \caption{Similar to Figure~\ref{fig:gmc7} for GMC 3}
    \label{fig:gmc3}
\end{figure*}
\end{appendix}

\end{document}